%% file: manuscript.tex
\PassOptionsToPackage{bookmarksopen=true,bookmarksnumbered=true,bookmarksdepth=2,pdfpagemode=UseOutlines}{hyperref}
\documentclass[a4paper,fleqn]{cas-sc}

\usepackage[numbers,sort]{natbib}
\usepackage{amsmath,amssymb}
\usepackage{float}
\usepackage{flafter}
\usepackage{placeins}


\ExplSyntaxOn
\cs_set:Npn \__make_tbl_caption:nn #1#2
{
  \l_tbl_align_tl
  \skip_vertical:N \l_tbl_abovecap_skip
  {\parbox{\dimexpr(\l_tbl_width_dim)}
    {\rightskip=0pt\rmfamily\small\textbf{\color{scolor}#1}\par#2\par\vskip4pt }}
  \skip_vertical:N \l_tbl_belowcap_skip
}
\cs_set:Npn \__make_fig_caption:nn #1#2
{
  \l_fig_align_tl
  \skip_vertical:N \l_fig_abovecap_skip
  \setbox\cascaptionbox=\hbox{
    \rmfamily\normalsize\textbf{\color{scolor}#1:}~#2
  }
  \ifdim\the\wd\cascaptionbox<\dim_use:N \l_fig_width_dim\relax
    \parbox{\l_fig_width_dim}
      {\unskip\ignorespaces\hfil\rmfamily\normalsize\textbf{\color{scolor}#1:}~#2\hfil\par}
  \else
    \parbox{\l_fig_width_dim}
      {\rightskip=0pt\unskip\ignorespaces\rmfamily\normalsize\textbf{\color{scolor}#1:}~#2\par}
  \fi
  \skip_vertical:N \l_fig_belowcap_skip
}
\ExplSyntaxOff

\begin{document}
\renewcommand{\topfraction}{0.9}
\renewcommand{\bottomfraction}{0.8}
\renewcommand{\textfraction}{0.08}
\renewcommand{\floatpagefraction}{0.8}
\setcounter{topnumber}{3}
\setcounter{bottomnumber}{2}
\setcounter{totalnumber}{5}

\shorttitle{Entropic lattice Boltzmann for anisotropic advection--diffusion}
\shortauthors{Feng et al.}

\title[mode=title]{Entropic lattice Boltzmann method for general anisotropic advection--diffusion}

\author{Jingsen Feng}
\author{Jing Leng}
\author{Jingchao Jiang}
\author{Xu Chu}
\cormark[1]
\ead{x.chu@exeter.ac.uk}

\affiliation{organization={Department of Engineering, University of Exeter},
  city={Exeter},
  postcode={EX4 4QF},
  country={United Kingdom}}

\cortext[cor1]{Corresponding author}

\begin{abstract}
Many transport processes exhibit direction-dependent diffusion, and their macroscopic description commonly leads to the full-tensor anisotropic advection--diffusion equation (ADE). This equation remains demanding for numerical discretizations when the principal axes are rotated relative to the mesh, since mixed derivatives and oblique fluxes can amplify grid-orientation errors under large tensor contrasts. This paper develops a local entropic lattice Boltzmann discretization for the general anisotropic advection--diffusion equation. The method separates the non-equilibrium population into a first-order flux sector and a residual ghost sector. The prescribed diffusion tensor is imposed directly through a local tensorial relaxation of the non-equilibrium flux, while the higher-order kinetic content is controlled by an ADE-corrected entropic stabilizer with a positivity fallback. Chapman--Enskog analysis shows that the scheme recovers the target full-tensor equation with the discrete-time diffusivity relation between the physical tensor and the flux-relaxation matrix. The resulting update is local and matrix-free, and applies to rotated, spatially varying, heterogeneous, and dynamically coupled tensor transport. The method is first validated using three-dimensional benchmarks, including advected Gaussian plumes, sinusoidal decay of rotated Fourier modes, and source-driven transport with spatially varying diffusion tensors. These tests examine off-diagonal tensor diffusion, high-P\'eclet-number advection, anisotropy ratios of $O(10^4)$, and local tensor contrasts up to $3\times10^4:1$. It is then applied to orientation-induced Taylor dispersion of elongated Brownian rods to quantify the dispersion enhancement induced by shear-driven rod rotation. The heat-conduction applications include rotated thermal-conductivity measurements and effective heat conduction in heterogeneous porous media, with anisotropy ratios up to $10^4:1$. Finally, anisotropic Rayleigh--B\'enard convection is simulated to assess buoyancy-coupled tensor transport and to examine how plume morphology and heat transfer change over seven decades of anisotropy ratios (from $10^{-4}$ to $10^3$). These results demonstrate that the proposed formulation provides an accurate and stable local solver for strongly anisotropic diffusion and advection--diffusion across diverse physical systems and engineering applications.
\end{abstract}


\maketitle

\input{sections/introduction}
\input{sections/methodology}
\input{sections/numerical}
\input{sections/application}
\FloatBarrier
\input{sections/conclusion}

\FloatBarrier
\clearpage
\appendix
\numberwithin{equation}{section}
\input{sections/appendix}

\bibliographystyle{unsrtnat}
\bibliography{cas-refs}

\end{document}

%% file: sections/introduction.tex
\section{Introduction}
\label{sec:introduction}

Directional scalar transport is a macroscopic signature of organized microstructure, material symmetry, and external forcing. When layers, fibers, pores, magnetic fields, or orientation distributions select preferred directions, scalar flux is governed by a tensor-valued mobility that couples magnitude and orientation. This tensor description appears in heat conduction and diffusion metamaterials \citep{Torquato2002,ZhangXuQuLeiLinOuyangJiangHuang2023NatRevPhys}, additively manufactured polymer and polymer-composite thermal materials \citep{ShemelyaEtAl2017AdditManuf,ButeEtAl2023IJAMT,CaiEtAl2022Polymers}, solute transport in porous and fractured media \citep{Bear1961JGR,PerezIllanesSoleMariFernandezGarcia2024AWR}, full-tensor flow and transport in petroleum reservoirs \citep{AavatsmarkEtAl1996JCP,AbushaikhaTerekhov2020JCP}, magnetized plasma and astrophysical heat transport \citep{Braginskii1965,BiriukovPrice2019MNRAS,GreenHuLoreMuStowell2022CPC}, biological tissue and cardiac electrophysiology \citep{BasserMattielloLeBihan1994BiophysJ,ZhangWangRezavandWuHu2021CMAME}, particulate suspensions \citep{Perrin1936,KumarEtAl2021}, and image analysis based on partial differential equations (PDEs) \citep{PeronaMalik1990PAMI,Paskas2025SignalProcessing}. Across these systems, the continuum description commonly reduces to an advection--diffusion equation (ADE)
\begin{equation}
\partial_t \phi+\nabla\cdot(\phi\mathbf{u})
=\nabla\cdot(\mathbf{D}\nabla\phi)+R,
\label{eq:intro-ade}
\end{equation}
where $\phi$ is a transported scalar, $\mathbf{u}$ is a prescribed or coupled velocity, $R$ is a source term, and $\mathbf{D}$ is a symmetric positive-definite diffusion tensor. The numerical difficulty lies in the tensorial operator. Writing $\mathbf{D}=\mathbf{Q}\boldsymbol{\Lambda}\mathbf{Q}^{\mathsf T}$, where $\mathbf{Q}$ contains the principal directions and $\boldsymbol{\Lambda}$ is the diagonal tensor of principal diffusivities, shows that a diagonal operator in the principal frame becomes a full operator in the computational frame when the principal axes are tilted. The off-diagonal entries generate mixed derivatives and oblique fluxes. Under strong anisotropy, grid misalignment can contaminate the weak perpendicular transport with errors from the dominant parallel flux, producing artificial transverse diffusion~\citep{VanEsKorenDeBlank2014JCP}. Related tensor--mesh effects also complicate discrete maximum-principle enforcement and can lead to loss of positivity or mesh-locking behavior~\citep{KuzminShashkovSvyatskiy2009JCP,ManziniPutti2007JCP}.

The finite-volume method (FVM), finite-element method (FEM), discontinuous Galerkin (DG), high-resolution finite-difference/finite-volume methods, and meshfree methods provide mature ways of solving diffusion and advection--diffusion equations. Their treatment of full anisotropic tensors is usually built around the construction of a stable and accurate numerical flux. For FVM, the two-point flux approximation is accurate only under restrictive alignment conditions, such as diffusion-tensor ($K$-) orthogonality between cell centers and faces~\citep{AavatsmarkEtAl1996JCP,TerekhovMallisonTchelepi2017JCP}. Full-tensor diffusion on general non-$K$-orthogonal grids has therefore motivated multi-point flux approximation (MPFA), mixed FVM, hybrid FVM, discrete duality FVM, mimetic finite-difference, and nonlinear monotone or extremum-preserving schemes~\citep{Droniou2014M3AS,DroniouEymardGallouetHerbin2010M3AS,HerbinHubert2008FVCA,TerekhovMallisonTchelepi2017JCP,DahmenDroniouRogier2022JCP}. These methods have strong conservation properties and broad mesh flexibility, but their computational structure can involve, depending on the formulation, interaction volumes, local reconstructions, auxiliary vertex or interface unknowns, nonlinear flux weights, and global algebraic systems. The same literature makes clear why the rotated, highly anisotropic case remains demanding. Flux-continuous MPFA variants may require M-matrix or quasi-M-matrix constructions, and high anisotropy can compromise monotonicity or formal accuracy \citep{VanEsKorenDeBlank2016JCP}. Nonlinear FVM schemes can recover positivity or minimum--maximum principles through nonlinear flux combinations, larger stencils, additional nonlinear iterations, or matrix-splitting conditions \citep{TerekhovMallisonTchelepi2017JCP,DahmenDroniouRogier2022JCP}.

The FEM and DG settings expose related algebraic issues. Standard Galerkin FEM formulations assemble a stiffness matrix whose off-diagonal signs and row sums govern discrete maximum principle behavior. For anisotropic diffusion, mesh-based sufficient conditions become difficult to satisfy; constrained FEM formulations adjust stiffness-matrix entries, split diffusive and antidiffusive algebraic fluxes, and solve nonlinear positivity-preserving correction problems \citep{KuzminShashkovSvyatskiy2009JCP}. DG and high-order FVM schemes offer high accuracy and geometric flexibility, yet diffusion fluxes must be defined across discontinuous polynomial traces. Interior-penalty, local DG, and diffusive generalized Riemann constructions address this interface problem by adding penalty terms, auxiliary gradient variables, or tailored diffusion fluxes \citep{ArnoldBrezziCockburnMarini2002SJNA,CockburnShu1998SJNA,GassnerLoercherMunz2007JCP}. Weighted essentially non-oscillatory (WENO)-type reconstructions are powerful for convection-dominated transport \citep{Shu2009SIAMReview}; when applied to convection--diffusion problems, however, the anisotropic diffusion term still requires a separate tensor-flux discretization, so monotonicity and stability are controlled primarily by the diffusion operator. Smoothed-particle hydrodynamics (SPH) removes the mesh, yet stable anisotropic diffusion remains delicate. Direct second-derivative SPH discretizations can become unstable for anisotropic heat conduction, and recent formulations therefore use two-first-derivative or complete-Hessian constructions to improve stability and suppress spurious oscillations or negative concentrations \citep{BiriukovPrice2019MNRAS,TangHuHaidn2026CMAME}. Field-aligned plasma heat transport is a useful extreme example because the parallel-to-perpendicular conductivity ratio may reach $10^{12}$, and non-aligned grids can transfer parallel truncation errors into spurious perpendicular transport \citep{VanEsKorenDeBlank2014JCP,VanEsKorenDeBlank2016JCP}. Finite-difference and finite-volume formulations have addressed such regimes by exploiting the magnetic-field tensor form $\mathbf{D}=(D_\parallel-D_\perp)\mathbf{b}\mathbf{b}^{\mathsf T}+D_\perp\mathbf{I}$, where $\mathbf{b}$ is the unit field direction and $D_\parallel$ and $D_\perp$ are the parallel and perpendicular diffusivities, together with field-line geometry and symmetric or high-order flux constructions, while high-order DG alternatives pair the resulting operators with auxiliary-space preconditioners for ill-conditioned linear systems \citep{VanEsKorenDeBlank2014JCP,VanEsKorenDeBlank2016JCP,GreenHuLoreMuStowell2022CPC}. These successes show that classical discretizations can handle selected extreme-anisotropy settings, but they do not amount to a general local ADE solver because robustness is usually tied to specialized tensor structure, tailored flux or connectivity choices, matrix-structure constraints, and preconditioned global algebraic solves. The general ADE in Eq.~\eqref{eq:intro-ade}, with arbitrary rotated, spatially varying tensors and coupled advection, remains a broader computational target.

Lattice Boltzmann (LB) methods offer a different route toward this broader ADE target. In scalar ADE-LB formulations, the transported field is represented by populations advanced through local collision and streaming, and the macroscopic diffusion law is recovered from the relaxation of non-equilibrium moments \citep{Ginzburg2005AWR,YoshidaNagaoka2010JCP,ChaiZhao2013PRE,ChaiShiGuo2016JSC}. An anisotropic LB scheme must therefore encode the physical tensor in the low-order flux sector while damping the remaining kinetic modes strongly enough that oblique, high-contrast tensors do not trigger instability or excessive numerical diffusion.

Several anisotropic ADE LB formulations have established the basic tensor-recovery mechanism. Equilibrium-type and link-type models were introduced for generic anisotropic dispersion \citep{Ginzburg2005AWR}. Multiple-relaxation-time (MRT) formulations then encoded rotated diffusion tensors through non-diagonal relaxation blocks in moment space \citep{YoshidaNagaoka2010JCP}, with later modifications of the relaxation matrix and equilibrium moments designed to remove unwanted macroscopic deviation terms \citep{HuangWu2014JCP}. Related work on convection--diffusion LB schemes clarified local non-equilibrium flux evaluation and extended MRT ideas to nonlinear anisotropic transport classes \citep{ChaiZhao2013PRE,ChaiShiGuo2016JSC,ZhaoWuChaiShi2020CAMWA}. Other developments have analyzed anisotropic collision/equilibrium parametrizations, diffusion-velocity or source-formulation routes, and local treatments of the convection term \citep{Ginzburg2013AWR,Perko2018ComputGeosci,HamilaJemniPerre2023IJP,LiLiXu2024CAMWA}. These studies show that full tensor ADEs can be recovered within LB moment dynamics.

The unresolved difficulty is the extreme rotated-tensor regime. Conventional MRT anisotropic ADE models embed the diffusion tensor in a full moment-space relaxation matrix, where a first-order block sets the leading Fickian tensor and the remaining non-conserved modes are assigned relaxation rates that affect truncation errors, boundary slip, and stability \citep{YoshidaNagaoka2010JCP,HuangWu2014JCP,ChaiShiGuo2016JSC}. This structure has proved effective for tensor recovery, but the rotated or fully anisotropic validation cases in representative MRT and block-relaxation studies remain at moderate tensor contrasts, typically of $O(10)$ or below; later source-formulation tests have emphasized removal of deviation terms and locality rather than extreme tensor contrast \citep{YoshidaNagaoka2010JCP,ChaiShiGuo2016JSC,ZhaoWuChaiShi2020CAMWA,LiLiXu2024CAMWA}. Diffusion-velocity single-relaxation-time (SRT) formulations form a useful exception for anisotropic hydrodynamic dispersion because they keep the SRT collision unchanged by converting the non-reference diffusive flux into an additional velocity inserted in the equilibrium distribution \citep{Perko2018ComputGeosci,PerkoPatel2014PRE}. For full tensors with off-diagonal entries, however, this diffusion velocity is reconstructed from a local system for the directional fluxes, so the tensor action is introduced through a macroscopic flux rewrite rather than through kinetic relaxation itself. These observations leave open a local kinetic construction in which a rotated tensor is imposed directly through the physical flux relaxation while the remaining kinetic content is stabilized separately.

This paper develops an entropic lattice Boltzmann method (ELBM) for anisotropic ADEs based on a flux-space separation. The physical operation is applied directly to the first-order non-equilibrium flux through a local $d\times d$ tensorial matrix that relaxes the flux and represents the prescribed diffusion tensor through the usual discrete-time LB diffusivity relation derived by Chapman--Enskog analysis. The residual higher-order population content is treated as a ghost sector controlled by a separate entropic mechanism. Cross-diffusion terms are carried by the same matrix-vector action that relaxes the physical flux, and the leading-order Fickian tensor is set by this flux relaxation alone. At each lattice node, the solver forms the non-equilibrium flux, applies the tensorial relaxation, embeds the relaxed flux back into population space, damps the ghost content using an ADE-corrected entropic amplitude, and streams. The scalar transport step avoids global matrix assembly and global linear or nonlinear solves. The resulting local kinetic update targets the general ADE in Eq.~\eqref{eq:intro-ade}, including arbitrary rotated tensors, spatially varying coefficients, coupled advection, and anisotropy ratios in the $O(10^4)$ range.

The entropic part follows the Karlin--B{\"o}sch--Chikatamarla (KBC) idea of controlling non-hydrodynamic content through an entropy-based stabilizer \citep{KarlinBoeschChikatamarla2014PRE,BoeschChikatamarlaKarlin2015PRE}. For scalar ADE, the usual hydrodynamic KBC stabilizer must be adjusted because the scalar equilibrium contains second-order velocity terms while the model has no hydrodynamic stress sector. We derive an ADE-corrected ghost amplitude that accounts for the overlap between the scalar equilibrium and the ghost subspace. The tensorial flux relaxation determines the macroscopic diffusion tensor, and the entropic amplitude damps the residual kinetic content on the positive branch. A geometric positivity fallback shortens the complete collision increment when needed, preserving the transported scalar exactly.

The method is verified and exercised over a set of increasingly demanding tests. A three-dimensional advected Gaussian plume measures the recovery of the full rotated diffusion tensor, including off-diagonal components, over anisotropy ratios up to $10^4:10:1$. A periodic sinusoidal-decay test isolates the projected decay rate $\mathbf{k}^{\mathsf T}\mathbf{D}\mathbf{k}$ for rotated constant tensors and keeps the relative decay-rate error at the level of $10^{-3}$ across the tested orientations. A source-driven three-dimensional benchmark with spatially varying coefficients tests high-P\'eclet-number anisotropic advection--diffusion at P\'eclet number $Pe=10^6$ and local anisotropy up to $3\times10^4:1$. The applications then move beyond manufactured solutions and include Taylor dispersion of elongated Brownian rods with orientation-induced cross-diffusion, effective thermal-conductivity measurement in rotated anisotropic solids and porous cube arrays, and anisotropic Rayleigh--B\'enard convection where the scalar solver is coupled to buoyancy-driven flow. Together these cases test the same local flux-space collision over constant, rotated, variable, steady, unsteady, source-driven, and dynamically coupled tensor-transport regimes.

The remainder of the paper is organized as follows. Section~\ref{sec:methodology} presents the lattice representation, flux/ghost decomposition, tensorial flux relaxation, entropic stabilizer, and positivity fallback. Section~\ref{sec:numerical} validates the method against analytical three-dimensional tensor-transport benchmarks. Section~\ref{sec:application} applies the solver to Brownian-rod dispersion, anisotropic thermal-conductivity measurements, porous anisotropic conduction, and anisotropic thermal convection. Appendix~\ref{app:ce-anisotropic-diffusion} gives the Chapman--Enskog derivation of the anisotropic ADE closure, and Appendix~\ref{app:ade-corrected-stabilizer} derives the ADE-corrected entropic ghost amplitude.

%% file: sections/methodology.tex
\section{Methodology}
\label{sec:methodology}

The method is constructed for the source-free transport operator associated with Eq.~\eqref{eq:intro-ade}, namely
\begin{equation*}
\partial_t \phi+\nabla\cdot(\phi\mathbf{u})
=\nabla\cdot(\mathbf{D}\nabla\phi),
\end{equation*}
where $\phi$, $\mathbf{u}$, and $\mathbf{D}$ have the same meanings as in Eq.~\eqref{eq:intro-ade}. The tensor may be constant or evaluated locally when the coefficients vary in space. The source terms and boundary conditions used in particular benchmarks are specified together with those cases; the collision operator below describes the transport step. We first define the lattice representation and the equilibrium moments, then separate the non-equilibrium population into a physical flux sector and a residual ghost sector. The diffusion tensor is imposed through a local tensorial relaxation of the flux, while the ghost content is controlled by an entropy-based scalar relaxation and a positivity fallback.

\subsection{Lattice Boltzmann formulation}

The scalar field is represented by populations $g_i(\mathbf{x},t)$ associated with a discrete velocity set $\{\mathbf{c}_i\}_{i=0}^{Q-1}$, following the standard lattice Boltzmann construction based on low-order discrete velocities~\citep{KarlinBoeschChikatamarla2014PRE,BoeschChikatamarlaKarlin2015PRE,QianDHumieresLallemand1992EPL}. The present formulation is implemented on tensor-product two-dimensional nine-velocity (D2Q9) and three-dimensional 27-velocity (D3Q27) lattices. In both cases,
\begin{equation}
\mathbf{c}_i\in\{-c,0,c\}^d,\qquad d\in\{2,3\},\qquad c=\frac{\Delta x}{\Delta t}.
\end{equation}
The one-dimensional quadrature weights are $w(0)=2/3$ and $w(\pm c)=1/6$. If
\begin{equation}
 m_i=\#\{\alpha\in\{1,\dots,d\}: |c_{i\alpha}|=c\},
\end{equation}
denotes the number of nonzero Cartesian components of $\mathbf{c}_i$, then the tensor-product weights are
\begin{equation}
W_i=\left(\frac{2}{3}\right)^{d-m_i}\left(\frac{1}{6}\right)^{m_i},\qquad c_s^2=\frac{c^2}{3}.
\end{equation}
Thus D2Q9 has weights $4/9$, $1/9$, and $1/36$ for the rest, axis, and diagonal velocities, while D3Q27 has weights $8/27$, $2/27$, $1/54$, and $1/216$ for velocities with $m_i=0$, $1$, $2$, and $3$. These tensor-product lattices provide the low-order isotropy needed by the advection--diffusion closure and leave enough non-hydrodynamic degrees of freedom to separate the first-order transport flux from higher-order ghost modes.

The populations satisfy
\begin{equation}
g_i(\mathbf{x}+\mathbf{c}_i\Delta t,t+\Delta t)-g_i(\mathbf{x},t)=\Omega_i(\mathbf{x},t),
\qquad i=0,\dots,Q-1.
\label{eq:method-lbe}
\end{equation}
The transported scalar is the zeroth moment,
\begin{equation}
\phi(\mathbf{x},t)=\sum_{i=0}^{Q-1}g_i.
\end{equation}
The prescribed velocity enters through the equilibrium distribution, which is chosen as the second-order low-Mach polynomial
\begin{equation}
g_i^{eq}=\phi W_i\left(
1+\frac{\mathbf{c}_i\cdot\mathbf{u}}{c_s^2}
+\frac{(\mathbf{c}_i\cdot\mathbf{u})^2}{2c_s^4}
-\frac{\mathbf{u}\cdot\mathbf{u}}{2c_s^2}
\right).
\label{eq:method-geq}
\end{equation}
It satisfies
\begin{equation}
\sum_i g_i^{eq}=\phi,\qquad
\sum_i\mathbf{c}_i g_i^{eq}=\phi\mathbf{u}.
\end{equation}
The first moment of the full population is therefore decomposed as
\begin{equation}
\sum_i\mathbf{c}_i g_i=\phi\mathbf{u}+\mathbf{j},
\qquad
\mathbf{j}=\sum_i\mathbf{c}_i\left(g_i-g_i^{eq}\right),
\label{eq:method-first-moment}
\end{equation}
where $\mathbf{j}$ is the first-order non-equilibrium flux. The quadratic velocity terms in \eqref{eq:method-geq} do not change the leading diffusive closure, but they reduce velocity-dependent truncation errors in advective transport and also enter the ADE-specific entropy correction introduced below.

\subsection{Flux/ghost decomposition}

Let
\begin{equation}
g_i^{neq}=g_i-g_i^{eq}.
\end{equation}
The non-equilibrium population is decomposed into a flux-carrying part and a higher-order ghost part,
\begin{equation}
g_i^{neq}=\Delta s_i+\Delta h_i.
\end{equation}
The flux $\mathbf{j}$ is a $d$-component vector, while the collision is applied to the $Q$ population components. We therefore need a population-space representative of a prescribed flux vector. For any vector $\mathbf{a}$, we use the first-order tensor-product quadrature lift
\begin{equation}
\Delta s_i=\mathcal{P}_i(\mathbf{j}),\qquad
\mathcal{P}_i(\mathbf{a})=\frac{W_i}{c_s^2}\mathbf{c}_i\cdot\mathbf{a},
\label{eq:method-lifting}
\end{equation}
which is the linear population perturbation whose only nonzero low-order moment is the prescribed first moment. Indeed, the quadrature identities give
\begin{equation}
\sum_i \mathcal{P}_i(\mathbf{a})
=\frac{1}{c_s^2}\sum_i W_i\,\mathbf{c}_i\cdot\mathbf{a}=0,
\end{equation}
and
\begin{equation}
\sum_i \mathbf{c}_i\mathcal{P}_i(\mathbf{a})
=\frac{1}{c_s^2}\sum_i W_i\,\mathbf{c}_i\mathbf{c}_i^{\top}\mathbf{a}
=\mathbf{a}.
\end{equation}
The factor $1/c_s^2$ is therefore the normalization required by the second-order lattice moment. Setting $\mathbf{a}=\mathbf{j}$ gives a population vector $\Delta s_i$ that carries exactly the non-equilibrium flux and no scalar mass. The ghost part is then defined as the residual
\begin{equation}
\Delta h_i=g_i^{neq}-\Delta s_i.
\end{equation}
By construction,
\begin{equation}
\sum_i\Delta s_i=\sum_i\Delta h_i=0,\qquad
\sum_i\mathbf{c}_i\Delta h_i=\mathbf{0},\qquad
\sum_i\mathbf{c}_i\Delta s_i=\mathbf{j}.
\end{equation}
Thus $\Delta s_i$ is the first-order part of $g_i^{neq}$, while $\Delta h_i$ contains the remaining kinetic content with no mass and no first-order flux. This separation is what allows the physical diffusion tensor to be assigned in flux space without prescribing a full moment-basis relaxation matrix.

\subsection{Tensorial relaxation of the diffusive flux}
\label{sec:method-collision}

The transport part of the collision relaxes the non-equilibrium flux through a matrix $\mathbf{S}\in\mathbb{R}^{d\times d}$:
\begin{equation}
\Omega_i^{(s)}=-\mathcal{P}_i(\mathbf{S}\mathbf{j}),\qquad
\mathbf{j}^{\star}=(\mathbf{I}-\mathbf{S})\mathbf{j}.
\label{eq:method-flux-relaxation}
\end{equation}
The matrix $\mathbf{S}$ is chosen from the target diffusion tensor by
\begin{equation}
\mathbf{S}=\left(\frac{1}{2}\mathbf{I}+\frac{\mathbf{D}}{c_s^2\Delta t}\right)^{-1},
\label{eq:method-s-from-d}
\end{equation}
or equivalently,
\begin{equation}
\mathbf{D}=c_s^2\Delta t\left(\mathbf{S}^{-1}-\frac{1}{2}\mathbf{I}\right).
\label{eq:method-d-from-s}
\end{equation}
The inverse relation shows where the discrete-time correction enters: the first-order Chapman--Enskog balance gives a tensorial Fick law proportional to $c_s^2\Delta t\,\mathbf{S}^{-1}$, while the second-order Taylor term of the lattice update subtracts the standard half-time-step contribution. Appendix~\ref{app:ce-anisotropic-diffusion} gives the corresponding multiscale derivation.

For a symmetric positive-definite tensor
\begin{equation}
\mathbf{D}=\mathbf{Q}\boldsymbol{\Lambda}\mathbf{Q}^{\top},\qquad
\boldsymbol{\Lambda}=\mathrm{diag}(D_1,\dots,D_d),\qquad D_k>0,
\end{equation}
the relaxation matrix has the same principal axes:
\begin{equation}
\mathbf{S}=\mathbf{Q}\,\mathrm{diag}(s_1,\dots,s_d)\,\mathbf{Q}^{\top},\qquad
s_k=\left(\frac{1}{2}+\frac{D_k}{c_s^2\Delta t}\right)^{-1}.
\label{eq:method-s-eigen}
\end{equation}
Hence $0<s_k<2$. Equations~\eqref{eq:method-s-from-d}--\eqref{eq:method-s-eigen} also give the local implementation route. If the material data are prescribed by principal diffusivities and orientation angles, the rotation matrix $\mathbf{Q}$ is built from those angles and the diagonal tensor $\boldsymbol{\Lambda}$ is formed directly. If the tensor is instead supplied in Cartesian components, a local eigendecomposition of the symmetric positive-definite matrix $\mathbf{D}$ provides the eigenvalues $D_k$ and eigenvectors in $\mathbf{Q}$. Degenerate eigenvalues do not create an ambiguity in $\mathbf{S}$, because the same scalar map $D_k\mapsto s_k$ is applied within the degenerate subspace.

The collision can then be viewed as a principal-frame flux update. The non-equilibrium flux is first projected onto the principal axes, each component is relaxed with its own scalar rate, and the result is rotated back:
\begin{equation}
\mathbf{j}_p=\mathbf{Q}^{\top}\mathbf{j},\qquad
\mathbf{j}_p^{\star}=\left[\mathbf{I}-\mathrm{diag}(s_1,\dots,s_d)\right]\mathbf{j}_p,\qquad
\mathbf{j}^{\star}=\mathbf{Q}\mathbf{j}_p^{\star}.
\label{eq:method-principal-flux-update}
\end{equation}
Equivalently, the collision uses the Cartesian action $\mathbf{S}\mathbf{j}=\mathbf{Q}\,\mathrm{diag}(s_1,\dots,s_d)\mathbf{Q}^{\top}\mathbf{j}$ in \eqref{eq:method-flux-relaxation}. The relaxed flux increment is then embedded back into population space through the lift $\mathcal{P}_i$. For spatially varying or dynamically coupled tensors, the same construction is applied independently at each lattice node and time step.

\subsection{Principal-axis interpretation and cross-diffusion}
\label{sec:method-cross-diffusion}

The tensorial relaxation above assigns diffusion through the full tensor $\mathbf{D}$, not through separate coordinate-direction diffusivities. This distinction matters when the principal axes of $\mathbf{D}$ are not aligned with the Cartesian lattice. The same tensor has a diagonal form in its principal coordinate frame and a generally non-diagonal form in the lattice coordinates. For a constant tensor, let $\boldsymbol{\xi}=\mathbf{Q}^{\top}\mathbf{x}$. Then
\begin{equation}
\nabla\cdot(\mathbf{D}\nabla\phi)
=\nabla_{\boldsymbol{\xi}}\cdot(\boldsymbol{\Lambda}\nabla_{\boldsymbol{\xi}}\phi)
=\sum_{k=1}^d D_k\,\partial_{\xi_k\xi_k}\phi.
\end{equation}
Thus diffusion is purely diagonal in the principal frame. When the same operator is written in the Cartesian coordinates used by the lattice, rotated principal axes appear as mixed-derivative terms.

In two dimensions,
\begin{equation}
\nabla\cdot(\mathbf{D}\nabla\phi)
=\partial_x(D_{xx}\partial_x\phi)
+\partial_x(D_{xy}\partial_y\phi)
+\partial_y(D_{yx}\partial_x\phi)
+\partial_y(D_{yy}\partial_y\phi).
\end{equation}
For a constant symmetric tensor, $D_{xy}=D_{yx}$ and
\begin{equation}
\nabla\cdot(\mathbf{D}\nabla\phi)
=D_{xx}\partial_{xx}\phi
+2D_{xy}\partial_{xy}\phi
+D_{yy}\partial_{yy}\phi.
\end{equation}

In three dimensions, a symmetric tensor has the Cartesian form
\begin{equation}
\mathbf{D}=
\begin{bmatrix}
D_{xx} & D_{xy} & D_{xz}\\
D_{xy} & D_{yy} & D_{yz}\\
D_{xz} & D_{yz} & D_{zz}
\end{bmatrix}.
\end{equation}
The corresponding divergence-form operator is
\begin{equation}
\begin{aligned}
\nabla\cdot(\mathbf{D}\nabla\phi)
={}&
\partial_x(D_{xx}\partial_x\phi)
+\partial_x(D_{xy}\partial_y\phi)
+\partial_x(D_{xz}\partial_z\phi)
\\
&+\partial_y(D_{xy}\partial_x\phi)
+\partial_y(D_{yy}\partial_y\phi)
+\partial_y(D_{yz}\partial_z\phi)
\\
&+\partial_z(D_{xz}\partial_x\phi)
+\partial_z(D_{yz}\partial_y\phi)
+\partial_z(D_{zz}\partial_z\phi).
\end{aligned}
\end{equation}
For constant symmetric $\mathbf{D}$, this reduces to
\begin{equation}
\begin{aligned}
\nabla\cdot(\mathbf{D}\nabla\phi)
={}&D_{xx}\partial_{xx}\phi
+D_{yy}\partial_{yy}\phi
+D_{zz}\partial_{zz}\phi
\\
&+2D_{xy}\partial_{xy}\phi
+2D_{xz}\partial_{xz}\phi
+2D_{yz}\partial_{yz}\phi .
\end{aligned}
\end{equation}
For spatially varying coefficients, the divergence form should be retained because derivatives of the tensor components also contribute to the macroscopic operator.

If the principal axes are tilted relative to the Cartesian lattice, the off-diagonal entries of $\mathbf{D}$ and $\mathbf{S}$ are the Cartesian-lattice representation of the same principal-axis diffusion tensor. Cross-diffusion is therefore produced by the local tensorial relaxation of the first-order flux, rather than by adding separate mixed-derivative stencils.

\subsection{Entropic stabilization with ADE correction}

The ghost sector is relaxed by a scalar entropic amplitude $\lambda$:
\begin{equation}
\Omega_i^{(h)}=-\lambda\Delta h_i.
\end{equation}
Combining the tensorial flux relaxation with the ghost relaxation gives the unrestricted collision operator
\begin{equation}
\Omega_i=\Omega_i^{(s)}+\Omega_i^{(h)}
=-\mathcal{P}_i(\mathbf{S}\mathbf{j})-\lambda\Delta h_i.
\label{eq:method-collision}
\end{equation}
In the isotropic limit $\mathbf{S}=2\beta\mathbf{I}$ with $\beta=(2\tau_{\mathrm{rel}})^{-1}$, where $\tau_{\mathrm{rel}}$ is the scalar relaxation time, this becomes the scalar ADE-KBC form
\begin{equation}
\Omega_i=-\beta\left(2\Delta s_i+\gamma\Delta h_i\right),\qquad \lambda=\beta\gamma.
\end{equation}
Compared with a conventional anisotropic MRT formulation, the present construction applies the non-diagonal first-order relaxation block directly to the physical flux. The higher-order modes are damped by the scalar entropic amplitude and do not change the leading-order tensor \eqref{eq:method-d-from-s}.

The entropy functional is
\begin{equation}
H[g]=\sum_i g_i\ln\left(\frac{g_i}{W_i}\right).
\label{eq:method-entropy}
\end{equation}
After the tensorial flux relaxation, but before the ghost relaxation, the intermediate state is
\begin{equation}
\widetilde{g}_i=g_i+\Omega_i^{(s)}
=g_i^{eq}+\Delta s_i^{\mathrm{post}}+\Delta h_i,
\end{equation}
where the superscript $\mathrm{post}$ refers only to the state after the flux relaxation,
\begin{equation}
\Delta s_i^{\mathrm{post}}
=\Delta s_i+\Omega_i^{(s)}
=\mathcal{P}_i\!\left[(\mathbf{I}-\mathbf{S})\mathbf{j}\right].
\end{equation}
The ghost update is taken along the line
\begin{equation}
g_i^{\star}=\widetilde{g}_i-\lambda\Delta h_i
=g_i^{eq}+\Delta s_i^{\mathrm{post}}+(1-\lambda)\Delta h_i.
\end{equation}
For positive trial populations, entropy stationarity in this ghost direction gives
\begin{equation}
\sum_i\Delta h_i
\ln\left(
\frac{g_i^{eq}+\Delta s_i^{\mathrm{post}}+(1-\lambda)\Delta h_i}{W_i}
\right)=0.
\label{eq:method-entropy-root}
\end{equation}

In all calculations reported in this paper, $\lambda$ is evaluated from the logarithmic expansion of \eqref{eq:method-entropy-root} about $g_i^{eq}$. Introducing
\begin{equation}
\langle x|y\rangle_H=\sum_i\frac{x_i y_i}{g_i^{eq}},
\end{equation}
the resulting entropic amplitude is
\begin{equation}
\lambda=1+
\frac{\langle\Delta h|\Delta s^{\mathrm{post}}\rangle_H+\mathcal{C}}
{\langle\Delta h|\Delta h\rangle_H},
\label{eq:method-lambda}
\end{equation}
with
\begin{equation}
\mathcal{C}=\sum_i\Delta h_i\ln\left(\frac{g_i^{eq}}{W_i}\right).
\label{eq:method-correction}
\end{equation}
If $\Delta h_i\equiv0$, the ghost correction is absent and we set $\lambda=1$. The term $\mathcal{C}$ is the ADE-specific correction associated with the overlap between the scalar ADE equilibrium and the ghost sector. It is retained because the quadratic velocity content of \eqref{eq:method-geq} is not represented by a separate hydrodynamic stress sector in the scalar transport model. Appendix~\ref{app:ade-corrected-stabilizer} derives \eqref{eq:method-lambda} from \eqref{eq:method-entropy-root}.

Anisotropy enters the stabilizer through $\Delta s^{\mathrm{post}}$, because the tensorial relaxation changes the flux-sector population before the ghost line is evaluated. The ghost relaxation itself remains scalar: $\mathbf{S}$ fixes the physical orientation and magnitude of the diffusive flux, while $\lambda$ damps the residual higher-order kinetic content.

\subsection{Positivity fallback}

The entropy evaluation requires positive populations. The implementation therefore includes a geometric fallback that shortens the complete collision increment if a trial update would leave the positive branch. Given a small threshold $\varepsilon>0$, define
\begin{equation}
\alpha=
\begin{cases}
1, & \Omega_i\ge 0 \quad \forall i,\\[4pt]
\min\left(1,\min\limits_{\Omega_i<0}\frac{g_i^n-\varepsilon}{-\Omega_i}\right), & \text{otherwise},
\end{cases}
\label{eq:method-alpha}
\end{equation}
and update
\begin{equation}
g_i^{n+1}=g_i^n+\alpha\Omega_i.
\label{eq:method-update}
\end{equation}
Because $\sum_i\Omega_i=0$, this fallback preserves the transported scalar exactly. When $\alpha=1$, the update is the entropic anisotropic collision operator \eqref{eq:method-collision}. When $\alpha<1$, both the flux relaxation and the ghost damping are under-relaxed together, so the Chapman--Enskog closure stated above applies to the positive smooth branch. In the numerical cases reported below, the entropy evaluation remains well defined on that branch.

\subsection{Algorithmic summary}

For one collision--streaming step:
\begin{enumerate}
\item Build $g_i^{eq}$ from \eqref{eq:method-geq}.
\item Evaluate the local tensor map \eqref{eq:method-s-from-d} and compute $g_i^{neq}$, $\mathbf{j}$, $\Delta s_i$, and $\Delta h_i$.
\item Apply the tensorial flux relaxation \eqref{eq:method-flux-relaxation} and form $\Delta s_i^{\mathrm{post}}$.
\item Evaluate the entropic amplitude from \eqref{eq:method-lambda}.
\item Form the collision increment \eqref{eq:method-collision}; apply \eqref{eq:method-alpha} only if the positivity fallback is needed.
\item Stream the post-collision populations according to \eqref{eq:method-lbe}.
\end{enumerate}

%% file: sections/numerical.tex
\section{Numerical Experiments}
\label{sec:numerical}

\subsection{Centroid-centered validation for an advected 3D Gaussian plume}
\label{sec:numerical-gaussian3d}

We begin with a three-dimensional Gaussian benchmark in which the diffusion tensor is intentionally tilted with respect to the Cartesian lattice. The governing equation is
\begin{equation}
\partial_t \phi+\mathbf{u}\cdot\nabla\phi=\nabla\cdot(\mathbf{D}\nabla\phi),
\end{equation}
posed on the periodic domain $[0,N_x)\times[0,N_y)\times[0,N_z)$ with $(N_x,N_y,N_z)=(256,256,256)$ and constant advection velocity $\mathbf{u}=(0.04,0.02,0.01)$. The initial condition is an isotropic Gaussian,
\begin{equation}
\phi(\mathbf{x},0)=\exp\!\left[-\frac{1}{2}(\mathbf{x}-\boldsymbol{\mu}_0)^{\mathsf{T}}\boldsymbol{\Sigma}_0^{-1}(\mathbf{x}-\boldsymbol{\mu}_0)\right],
\qquad
\boldsymbol{\mu}_0=\left(\frac{N_x-1}{2},\frac{N_y-1}{2},\frac{N_z-1}{2}\right),
\qquad
\boldsymbol{\Sigma}_0=\sigma_0^2\mathbf{I},
\end{equation}
with $\sigma_0=8$. For constant $\mathbf{u}$ and $\mathbf{D}$, the exact solution remains Gaussian,
\begin{equation}
\phi(\mathbf{x},t)=\sqrt{\frac{\det\boldsymbol{\Sigma}_0}{\det\boldsymbol{\Sigma}(t)}}\,
\exp\!\left[-\frac{1}{2}(\mathbf{x}-\boldsymbol{\mu}(t))^{\mathsf{T}}\boldsymbol{\Sigma}(t)^{-1}(\mathbf{x}-\boldsymbol{\mu}(t))\right],
\end{equation}
with
\begin{equation}
\boldsymbol{\mu}(t)=\boldsymbol{\mu}_0+\mathbf{u}t,
\qquad
\boldsymbol{\Sigma}(t)=\boldsymbol{\Sigma}_0+2\mathbf{D}t.
\end{equation}
In the periodic post-processing, the centroid is wrapped back into the computational box and the fields are recentered on $\boldsymbol{\mu}(t)$ so that the tensor-induced deformation can be examined without the bulk translation. All four cases share the same oblique orientation, represented here by the ZYZ Euler-angle convention
\begin{equation}
\mathbf{R}=\mathbf{R}_z(\alpha_E)\mathbf{R}_y(\beta_E)\mathbf{R}_z(\gamma_E),
\qquad
(\alpha_E,\beta_E,\gamma_E)=(120^\circ,90^\circ,135^\circ),
\end{equation}
where $\mathbf{R}_z$ and $\mathbf{R}_y$ denote standard rotations about the $z$- and $y$-axes. The diffusion tensor is then obtained from the principal-frame coefficients through $\mathbf{D}=\mathbf{R}\,\mathrm{diag}(d_1,d_2,d_3)\mathbf{R}^{\mathsf T}$. The validation therefore addresses the recovery of the principal diffusivities and the off-diagonal tensor coefficients in the same set of calculations. To probe the tensorial response over a wide range of anisotropy, we consider four sets of principal diffusivities,
\begin{equation}
(d_1,d_2,d_3)\in\left\{
(10^{-2},10^{-2},10^{-6}),\,
(10^{-2},10^{-3},10^{-6}),\,
(10^{-2},10^{-4},10^{-6}),\,
(10^{-2},10^{-5},10^{-6})
\right\},
\end{equation}
with the tilted principal frame held fixed. The discussion below is organized in terms of the normalized time
\begin{equation}
\tau=\frac{D_{\mathrm{ref}}t}{\sigma_0^2},
\qquad
D_{\mathrm{ref}}=d_1=10^{-2},
\end{equation}
as reported in Fig.~\ref{fig:gaussian3d_profiles}.

\begin{figure}[pos=H]
    \centering
    \includegraphics[width=0.98\textwidth]{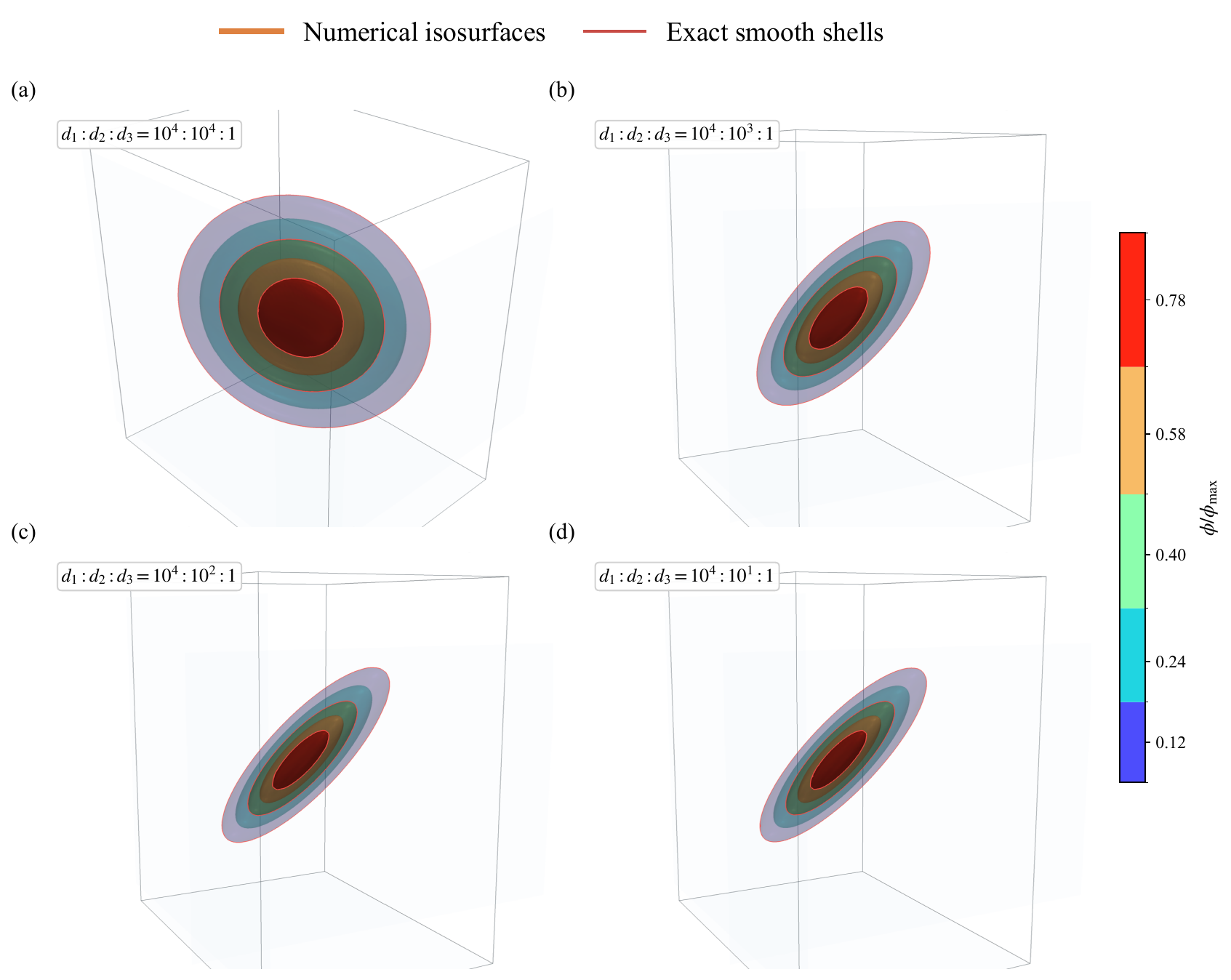}
    \caption{Centroid-centered three-dimensional concentration distributions at the largest normalized time shown in Fig.~\ref{fig:gaussian3d_profiles} ($\tau=6.25$) for four principal-diffusivity ratios. In each panel, five semi-transparent numerical isosurfaces are overlaid with smooth analytical shells extracted at the same normalized levels $\phi/\phi_{\max}$, reported by the color bar. This representation makes the agreement between the numerical and exact plume shapes directly visible in three dimensions. Panel (a) corresponds to $(d_1,d_2,d_3)=(10^{-2},10^{-2},10^{-6})$; panels (b)--(d) decrease the intermediate diffusivity from $10^{-3}$ to $10^{-5}$, with $d_1=10^{-2}$ and $d_3=10^{-6}$ unchanged.}
    \label{fig:gaussian3d_isosurfaces}
\end{figure}

Figure~\ref{fig:gaussian3d_isosurfaces} provides a direct three-dimensional comparison between the numerical and analytical plume shapes in the centroid-centered frame. In the nearly axisymmetric case of Fig.~\ref{fig:gaussian3d_isosurfaces}(a), where $d_1=d_2$, the scalar cloud spreads almost equally in the two fast principal directions. It remains tightly compressed along the slow direction and forms a flattened spheroidal structure. As $d_2$ is reduced from $10^{-3}$ to $10^{-5}$ in Figs.~\ref{fig:gaussian3d_isosurfaces}(b)--\ref{fig:gaussian3d_isosurfaces}(d), the in-plane symmetry is progressively broken and the plume evolves into a more elongated ellipsoid aligned with the dominant principal axis. With the tilted principal frame fixed, the dominant plume orientation is preserved as the anisotropy ratio changes, and the primary geometric response is a strong variation in aspect ratio. For case (a), the equality $d_1=d_2$ makes the in-plane orientation non-unique. The overlaid analytical shells closely track the numerical isosurfaces at all five levels, showing that the scheme recovers the overall orientation of the rotated tensor together with the full three-dimensional spreading and thickness of the Gaussian plume.

\begin{figure}[pos=H]
    \centering
    \includegraphics[width=0.98\textwidth]{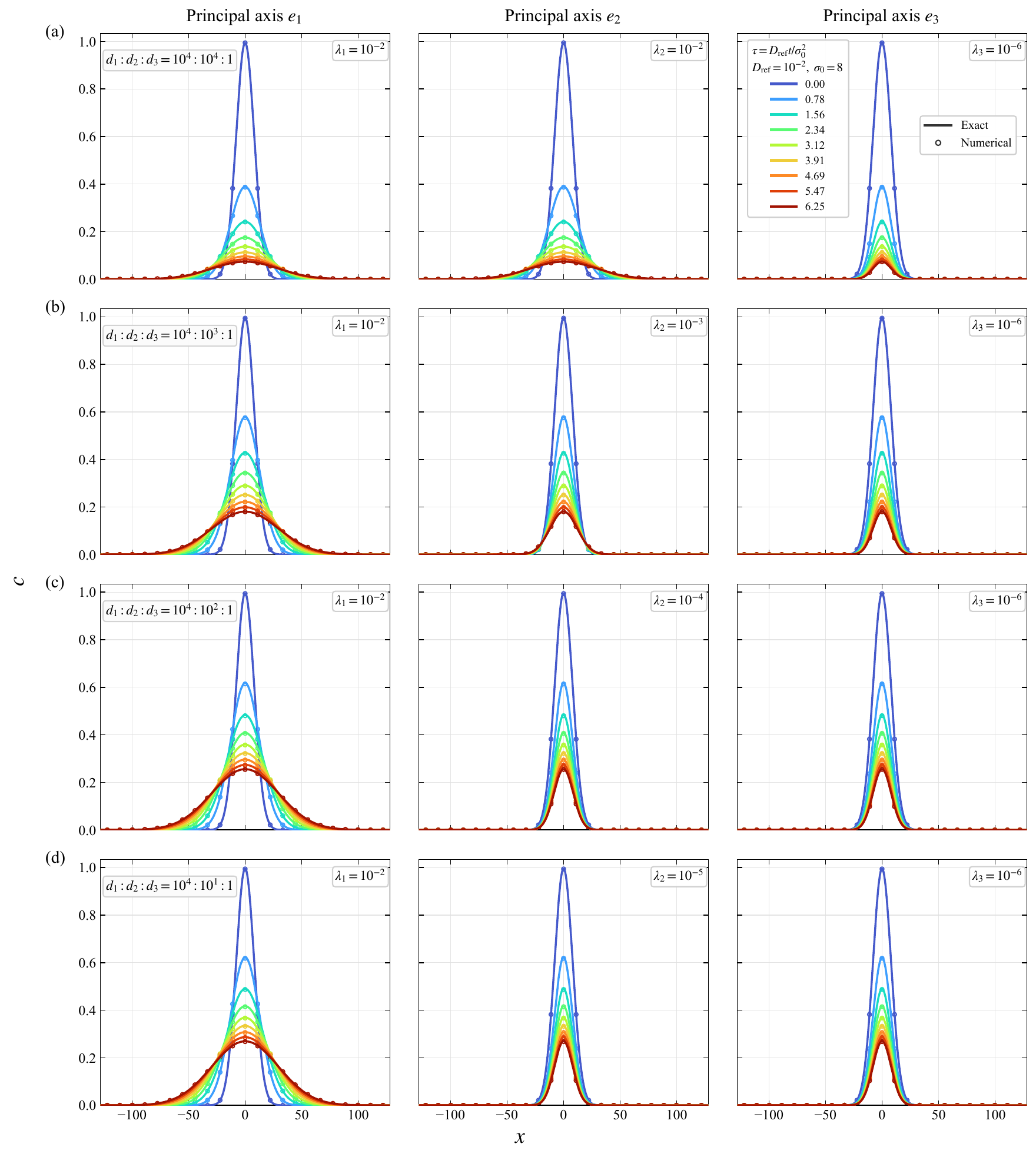}
    \caption{Evolution of centroid-centered concentration profiles along the three principal axes. Rows (a)--(d) correspond to the four diffusivity ratios in Fig.~\ref{fig:gaussian3d_isosurfaces}. Columns show cuts along the eigenvectors associated with the principal diffusivities $d_1$, $d_2$, and $d_3$, respectively. Colors denote the normalized time $\tau$ listed in the legend. Solid lines are the analytical Gaussian profiles and open circles are the numerical samples.}
    \label{fig:gaussian3d_profiles}
\end{figure}

Figure~\ref{fig:gaussian3d_profiles} gives a more detailed view of the same dynamics through one-dimensional cuts taken along the principal axes of the prescribed diffusion tensor. In the first row, the profiles along the first and second principal axes are almost indistinguishable because the two largest eigenvalues are equal. Once $d_2$ is reduced, the second column narrows in a controlled and monotone manner from row (b) to row (d). The first column remains governed by the fixed value $d_1=10^{-2}$, and the third column stays sharply localized because $d_3=10^{-6}$ is unchanged. The agreement between analytical curves and numerical samples is consistently tight over the whole time sequence, including the late-time profiles where the plume has spread over a much larger support. Small visible differences are confined to the far tails, where the concentration is already several orders of magnitude below the peak.

\begin{table}[pos=htbp]
    \centering
    \rmfamily
    \caption{Comparison between the prescribed diffusion tensor and the tensor reconstructed from the numerical second moments at the largest normalized time shown in Fig.~\ref{fig:gaussian3d_profiles}. For compactness, the six independent tensor coefficients are reported in units of $10^{-3}$.}
    \label{tab:gaussian3d_tensor}
    \begingroup
    \small
    \setlength{\tabcolsep}{2.6pt}
    \renewcommand{\arraystretch}{1.08}
    \begin{tabular*}{\textwidth}{@{\extracolsep{\fill}}llrrrrrr@{}}
        \toprule
        Case $(d_1:d_2:d_3)$ & Type & $10^3D_{xx}$ & $10^3D_{xy}$ & $10^3D_{xz}$ & $10^3D_{yy}$ & $10^3D_{yz}$ & $10^3D_{zz}$ \\
        \midrule
        \multirow{2}{*}{$10^4:10^4:1$} & Target    & 7.5003 & 4.3297 & 0.0000  & 2.5007 & 0.0000  & 10.0000 \\
                                         & Recovered & 7.5025 & 4.3296 & 0.0000  & 2.5029 & 0.0000  & 10.0016 \\
        \multirow{2}{*}{$10^4:10^3:1$} & Target    & 4.1253 & 2.3811 & -3.8971 & 1.3757 & -2.2500 & 5.5000  \\
                                         & Recovered & 4.1274 & 2.3812 & -3.8971 & 1.3778 & -2.2500 & 5.5021  \\
        \multirow{2}{*}{$10^4:10^2:1$} & Target    & 3.7878 & 2.1863 & -4.2868 & 1.2632 & -2.4750 & 5.0500  \\
                                         & Recovered & 3.7899 & 2.1863 & -4.2868 & 1.2653 & -2.4750 & 5.0521  \\
        \multirow{2}{*}{$10^4:10^1:1$} & Target    & 3.7540 & 2.1668 & -4.3258 & 1.2520 & -2.4975 & 5.0050  \\
                                         & Recovered & 3.7561 & 2.1668 & -4.3258 & 1.2541 & -2.4975 & 5.0071  \\
        \bottomrule
    \end{tabular*}
    \endgroup
\end{table}

Table~\ref{tab:gaussian3d_tensor} highlights the effect of the tensor rotation. The shared oblique orientation generates nonzero cross-diffusion coefficients, so the benchmark is governed by the full tensor, including the principal widths and the cross terms. In the first case, the equality $d_1=d_2$ removes the out-of-plane asymmetry and leaves only the oblique $xy$ coupling. Once this degeneracy is lifted, the same rotation yields substantial $D_{xz}$ and $D_{yz}$ entries. The recovered coefficients remain very close to the prescribed ones across all four cases, with differences appearing only in the fourth decimal place of the scaled values listed in the table.

The moment statistics confirm the visual impression. At the largest normalized time, the relative mass drift is below $2.2\times10^{-12}$ in all four runs, the centroid error remains below $5.9\times10^{-5}$ grid units, and the relative covariance error lies between $2.2\times10^{-4}$ and $3.4\times10^{-4}$. Reconstructing the diffusion tensor from the numerical second moments gives a Frobenius-norm error of about $3.4\times10^{-6}$ to $3.7\times10^{-6}$. These results show that the present ELBM formulation captures uniform advection, tensor rotation, and strongly anisotropic diffusion within a single update. The correct spreading rates and Gaussian profile evolution are preserved throughout the calculation.

\subsection{3D sinusoidal-decay validation for a rotated constant diffusion tensor}
\label{sec:numerical-decay3d}

We next consider a periodic three-dimensional decay benchmark designed to isolate the diffusive response of the rotated tensor without the additional effects of advection, spatially varying coefficients, or source forcing. The governing equation is
\begin{equation}
\partial_t \phi=\nabla\cdot(\mathbf{D}\nabla\phi),
\end{equation}
with a constant diffusion tensor $\mathbf{D}$ and periodic boundary conditions on a cubic domain of size $256^3$. The initial condition is prescribed as a single Fourier mode,
\begin{equation}
\phi(x,y,z,0)=c_0+A_0\cos(k_x x+k_y y+k_z z),
\end{equation}
where $c_0=1$ and $A_0=10^{-2}$. The wavevector is defined from the integer mode indices $(m_x,m_y,m_z)$ as
\begin{equation}
\mathbf{k}=\left(\frac{2\pi m_x}{N_x},\,\frac{2\pi m_y}{N_y},\,\frac{2\pi m_z}{N_z}\right),
\qquad
(N_x,N_y,N_z)=(256,256,256),
\end{equation}
so the mode indices enter the decay problem only through the projected rate $\alpha=\mathbf{k}^{\mathsf{T}}\mathbf{D}\mathbf{k}$. For a constant tensor, the exact solution preserves the modal shape and only the amplitude changes in time,
\begin{equation}
\phi(x,y,z,t)=c_0+A_0e^{-\alpha t}\cos(k_x x+k_y y+k_z z),
\qquad
\alpha=\mathbf{k}^{\mathsf{T}}\mathbf{D}\mathbf{k}.
\end{equation}
This case therefore provides a direct verification of the decay rate predicted by the prescribed rotated tensor.

For the main comparison we use the same four anisotropy ratios as in the Gaussian and source-driven benchmarks,
\begin{equation}
(d_1,d_2,d_3)\in\left\{
(10^{-2},10^{-2},10^{-6}),\,
(10^{-2},10^{-3},10^{-6}),\,
(10^{-2},10^{-4},10^{-6}),\,
(10^{-2},10^{-5},10^{-6})
\right\},
\end{equation}
and fix the principal-frame orientation at the ZYZ Euler angles $(\alpha_E,\beta_E,\gamma_E)=(120^\circ,90^\circ,135^\circ)$. The measured mode is $(m_x,m_y,m_z)=(3,3,0)$, which yields a robust oblique projection of the rotated tensor across all four anisotropy ratios. Each run is advanced to the common normalized time window $\tau=\alpha_{\mathrm{theory}}t\approx2$, so the expected amplitude reduction is close to $e^{-2}$. In post-processing, the modal amplitude is extracted from the discrete Fourier coefficient at $(m_x,m_y,m_z)$ and the ELBM decay rate $\alpha_{\mathrm{ELBM}}$ is obtained from a least-squares fit of $\ln A(t)$ against time.

\begin{figure}[pos=H]
    \centering
    \includegraphics[width=0.98\textwidth]{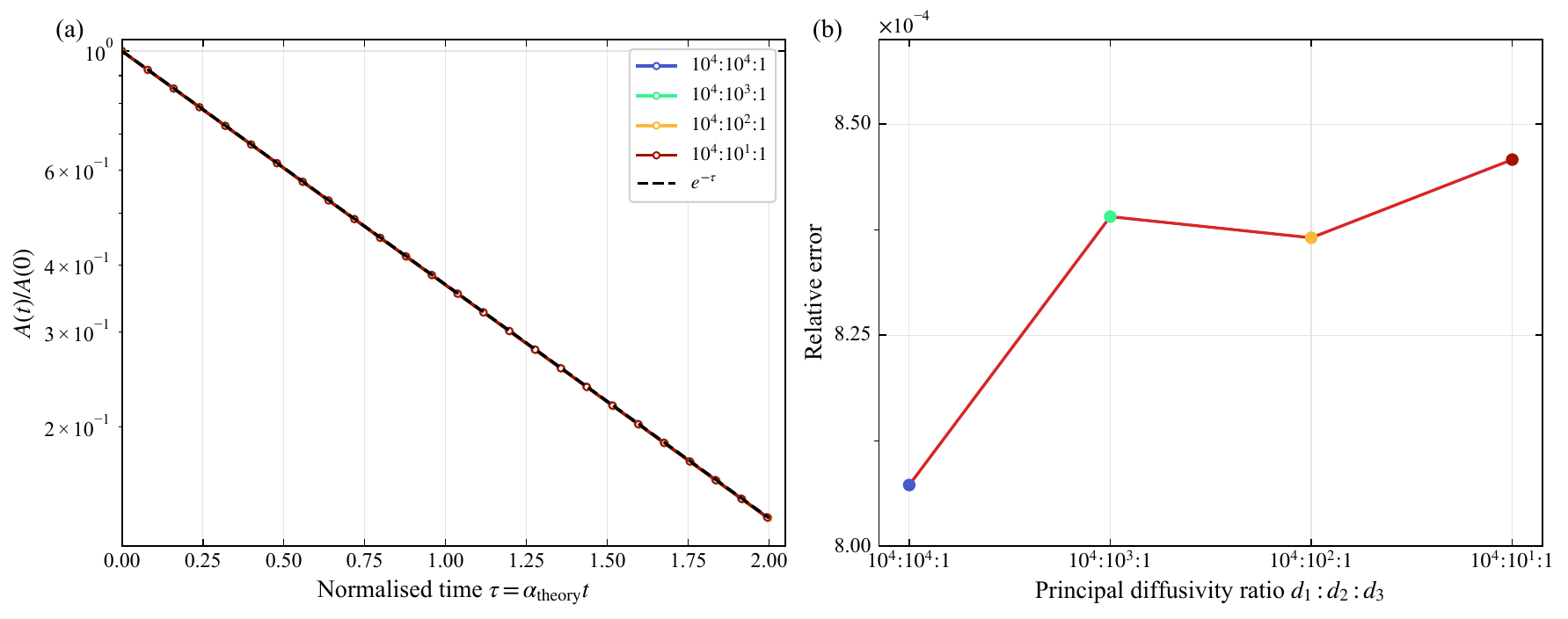}
    \caption{Sinusoidal-decay validation for the four rotated constant-tensor cases. Panel (a) shows the normalized modal amplitudes $A(t)/A(0)$ against the normalized time $\tau=\alpha_{\mathrm{theory}}t$ for the four anisotropy ratios $10^4\!:\!10^4\!:\!1$, $10^4\!:\!10^3\!:\!1$, $10^4\!:\!10^2\!:\!1$, and $10^4\!:\!10^1\!:\!1$. The dashed line denotes the exact decay law $e^{-\tau}$. Panel (b) reports the relative decay-rate error $\lvert \alpha_{\mathrm{ELBM}}-\alpha_{\mathrm{theory}}\rvert/\alpha_{\mathrm{theory}}$ for the same four cases.}
    \label{fig:decay3d_main}
\end{figure}

Figure~\ref{fig:decay3d_main}(a) shows that the four normalized decay curves collapse onto the analytical law $e^{-\tau}$ over the full observation window. This is a stringent test because the rotated tensor changes the effective diffusivity through the projected combination $\mathbf{k}^{\mathsf{T}}\mathbf{D}\mathbf{k}$, while the principal anisotropy ratio spans four orders of magnitude. The collapse indicates that the present formulation recovers the correct diffusive attenuation along the chosen oblique mode for all four tensors. The corresponding relative errors in Fig.~\ref{fig:decay3d_main}(b) lie in the narrow range from $8.07\times10^{-4}$ to $8.46\times10^{-4}$. The small variation across the four cases shows that the decay-rate accuracy is only weakly affected by the reduction of the intermediate principal diffusivity from $10^{-2}$ to $10^{-5}$.

To test the sensitivity to tensor orientation, we perform an additional Euler-angle sweep for the representative anisotropy ratio $(d_1,d_2,d_3)=(10^{-2},10^{-4},10^{-6})$, corresponding to $10^4\!:\!10^2\!:\!1$. The mode is kept at $(m_x,m_y,m_z)=(3,3,0)$, the angles $\alpha_E=120^\circ$ and $\gamma_E=135^\circ$ are fixed, and the out-of-plane angle is varied over $\beta_E\in\{0^\circ,30^\circ,60^\circ,90^\circ\}$. Each of these four runs is again advanced to $\tau\approx2$ before extracting $\alpha_{\mathrm{ELBM}}$ from the modal history.

\begin{figure}[pos=H]
    \centering
    \includegraphics[width=0.98\textwidth]{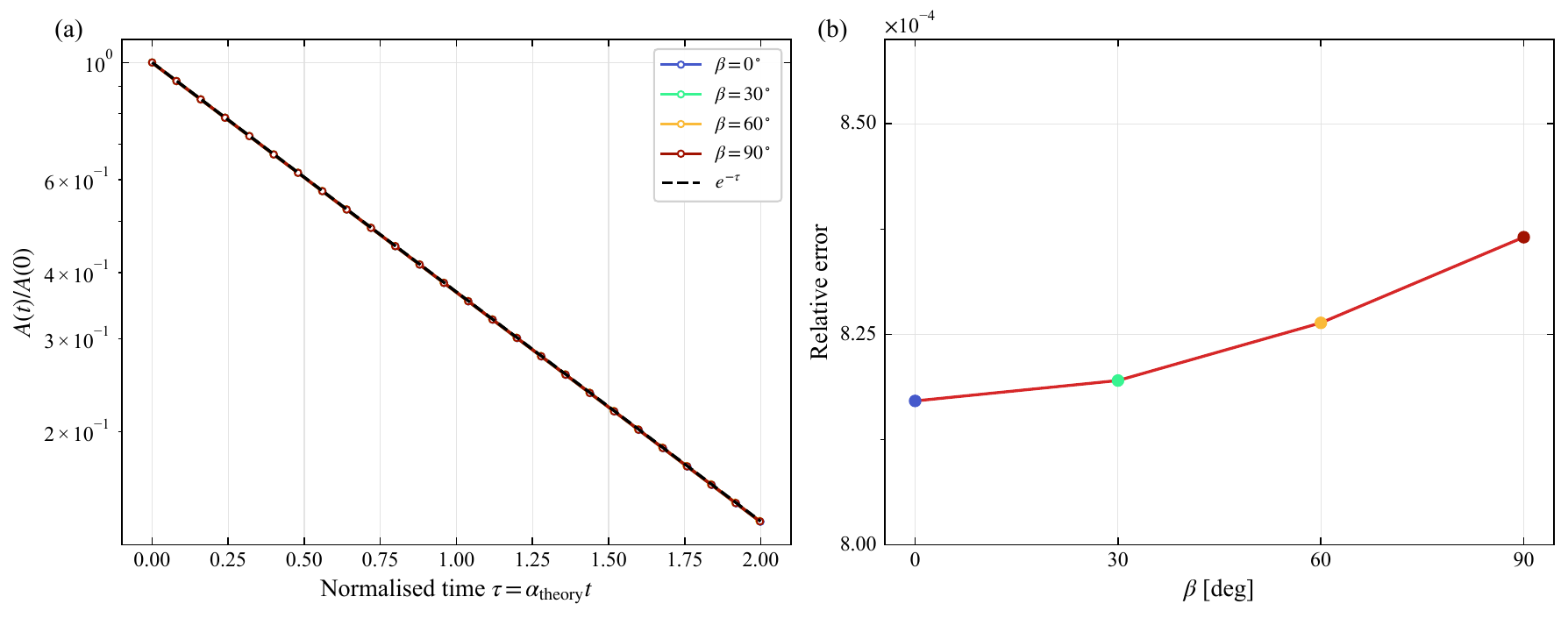}
    \caption{Sensitivity of the sinusoidal-decay benchmark to the Euler angle $\beta_E$ for the fixed anisotropy ratio $d_1:d_2:d_3=10^4\!:\!10^2\!:\!1$. Panel (a) presents the normalized decay curves for $\beta_E=0^\circ,30^\circ,60^\circ,$ and $90^\circ$ together with the exact law $e^{-\tau}$. Panel (b) shows the corresponding relative decay-rate errors.}
    \label{fig:decay3d_beta}
\end{figure}

Figure~\ref{fig:decay3d_beta}(a) confirms that the normalized decay curves remain collapsed when the principal frame is tilted out of plane. The measured error levels in Fig.~\ref{fig:decay3d_beta}(b) remain between $8.17\times10^{-4}$ and $8.37\times10^{-4}$ over the entire $\beta_E$ sweep. The weak dependence on $\beta_E$ indicates that the decay-rate accuracy is governed by the resolved modal projection and remains essentially insensitive to the orientation of the Cartesian lattice. In combination with the Gaussian benchmark above, this result supports the conclusion that the present ELBM recovers both the global tensor geometry and the projected constant-tensor diffusion rate in three dimensions.

\begin{table}[pos=htbp]
    \centering
    \rmfamily
    \caption{Summary of the decay-rate verification for the rotated constant-tensor benchmark. The upper block corresponds to the anisotropy-ratio sweep at fixed Euler angles $(\alpha_E,\beta_E,\gamma_E)=(120^\circ,90^\circ,135^\circ)$; the lower block reports the $\beta_E$ sweep at the fixed anisotropy ratio $10^4\!:\!10^2\!:\!1$.}
    \label{tab:decay3d_rates}
    \begingroup
    \small
    \setlength{\tabcolsep}{2.6pt}
    \renewcommand{\arraystretch}{1.08}
    \newcommand{\ratecell}[1]{\makebox[2.35cm][c]{$#1$}}
    \newcommand{\errcell}[1]{\makebox[1.85cm][c]{$#1$}}
    \begin{tabular*}{\textwidth}{@{\extracolsep{\fill}}lccccc@{}}
        \toprule
        Case & $(\alpha_E,\beta_E,\gamma_E)$ & $(m_x,m_y,m_z)$ & \makebox[2.35cm][c]{$\alpha_{\mathrm{theory}}$} & \makebox[2.35cm][c]{$\alpha_{\mathrm{ELBM}}$} & \makebox[1.85cm][c]{Relative error} \\
        \midrule
        \multicolumn{6}{c}{Anisotropy-ratio sweep at fixed Euler angles} \\
        \midrule
        $10^4:10^4:1$ & $(120^\circ,90^\circ,135^\circ)$ & $(3,3,0)$ & \ratecell{1.01168\times10^{-4}} & \ratecell{1.01250\times10^{-4}} & \errcell{8.07\times10^{-4}} \\
        $10^4:10^3:1$ & $(120^\circ,90^\circ,135^\circ)$ & $(3,3,0)$ & \ratecell{5.56427\times10^{-5}} & \ratecell{5.56894\times10^{-5}} & \errcell{8.39\times10^{-4}} \\
        $10^4:10^2:1$ & $(120^\circ,90^\circ,135^\circ)$ & $(3,3,0)$ & \ratecell{5.10902\times10^{-5}} & \ratecell{5.11329\times10^{-5}} & \errcell{8.37\times10^{-4}} \\
        $10^4:10^1:1$ & $(120^\circ,90^\circ,135^\circ)$ & $(3,3,0)$ & \ratecell{5.06349\times10^{-5}} & \ratecell{5.06778\times10^{-5}} & \errcell{8.46\times10^{-4}} \\
        \midrule
        \multicolumn{6}{c}{$\beta_E$ sweep at fixed anisotropy ratio $10^4:10^2:1$} \\
        \midrule
        $\beta_E=0^\circ$  & $(120^\circ,0^\circ,135^\circ)$  & $(3,3,0)$ & \ratecell{8.15941\times10^{-5}} & \ratecell{8.16608\times10^{-5}} & \errcell{8.17\times10^{-4}} \\
        $\beta_E=30^\circ$ & $(120^\circ,30^\circ,135^\circ)$ & $(3,3,0)$ & \ratecell{7.70819\times10^{-5}} & \ratecell{7.71450\times10^{-5}} & \errcell{8.20\times10^{-4}} \\
        $\beta_E=60^\circ$ & $(120^\circ,60^\circ,135^\circ)$ & $(3,3,0)$ & \ratecell{6.54253\times10^{-5}} & \ratecell{6.54794\times10^{-5}} & \errcell{8.26\times10^{-4}} \\
        $\beta_E=90^\circ$ & $(120^\circ,90^\circ,135^\circ)$ & $(3,3,0)$ & \ratecell{5.10902\times10^{-5}} & \ratecell{5.11329\times10^{-5}} & \errcell{8.37\times10^{-4}} \\
        \bottomrule
    \end{tabular*}
    \endgroup
\end{table}

Table~\ref{tab:decay3d_rates} summarizes the fitted rates for both the anisotropy-ratio sweep at fixed Euler angles and the $\beta_E$ sweep at fixed anisotropy ratio. The agreement between $\alpha_{\mathrm{ELBM}}$ and $\alpha_{\mathrm{theory}}$ is uniformly close, and the relative error stays at the level of $8\times10^{-4}$ throughout. This benchmark therefore provides a clean spectral verification of the constant-tensor operator before the introduction of spatially varying coefficients and source forcing in the next subsection.

\FloatBarrier

\subsection{3D anisotropic convection--diffusion equation with constant velocity and variable diffusion tensor}
\label{sec:numerical-chai3d}

We next consider the three-dimensional anisotropic advection--diffusion equation with a constant velocity $\mathbf{u}=(u_x,u_y,u_z)=(1,1,1)$ and a variable diffusion tensor $\mathbf{K}$,
\begin{equation}
\partial_t \phi+\nabla\cdot(\phi\mathbf{u})=\nabla\cdot(\mathbf{K}\nabla\phi)+R,
\end{equation}
where $R$ is the source term. This problem is more demanding than the constant-coefficient case because $\mathbf{K}$ depends on space and the governing equation cannot, in general, be reduced to an equivalent isotropic form. In the present three-dimensional extension, the diffusion tensor is taken as
\begin{equation}
\mathbf{K}(x,y,z)=\mathrm{diag}\!\left(d_1\left[2-\sin(2\pi x)\sin(2\pi y)\sin(2\pi z)\right],\,d_2,\,d_3\right),
\end{equation}
with $d_1$, $d_2$, and $d_3$ denoting the baseline principal diffusivities. The source term is prescribed as
\begin{equation}
\begin{aligned}
R={}&\exp(At)\Bigl\{
\sin(2\pi x)\sin(2\pi y)\sin(2\pi z)
\\
&\quad +4d_1\pi^2\cos(4\pi x)\sin^2(2\pi y)\sin^2(2\pi z)
\\
&\quad +2\pi\bigl[
u_x\cos(2\pi x)\sin(2\pi y)\sin(2\pi z)
+u_y\sin(2\pi x)\cos(2\pi y)\sin(2\pi z)
+u_z\sin(2\pi x)\sin(2\pi y)\cos(2\pi z)
\bigr]
\Bigr\},
\end{aligned}
\end{equation}
where
\begin{equation}
A=1-4\pi^2(2d_1+d_2+d_3).
\end{equation}
Under periodic boundary conditions on the physical region $[0,1)^3$ and the initial condition
\begin{equation}
\phi(x,y,z,0)=\sin(2\pi x)\sin(2\pi y)\sin(2\pi z),
\end{equation}
the exact solution of the problem is
\begin{equation}
\phi(x,y,z,t)=\exp(At)\sin(2\pi x)\sin(2\pi y)\sin(2\pi z).
\end{equation}
Since
\begin{equation}
K_{xx}\in[d_1,3d_1],
\end{equation}
the local anisotropy is stronger than the baseline ratio in part of the domain. The ratios reported below denote the baseline coefficients $(d_1,d_2,d_3)$. In the labeled cases, $d_1:d_3=10^4:1$. Since $K_{xx}$ varies from $d_1$ to $3d_1$, the pointwise principal ratio $K_{xx}:K_{zz}$ reaches $3\times10^4:1$, and the maximum local anisotropy is $3\times10^4:1$.

\begin{figure}[pos=htbp]
    \centering
    \includegraphics[width=0.92\textwidth]{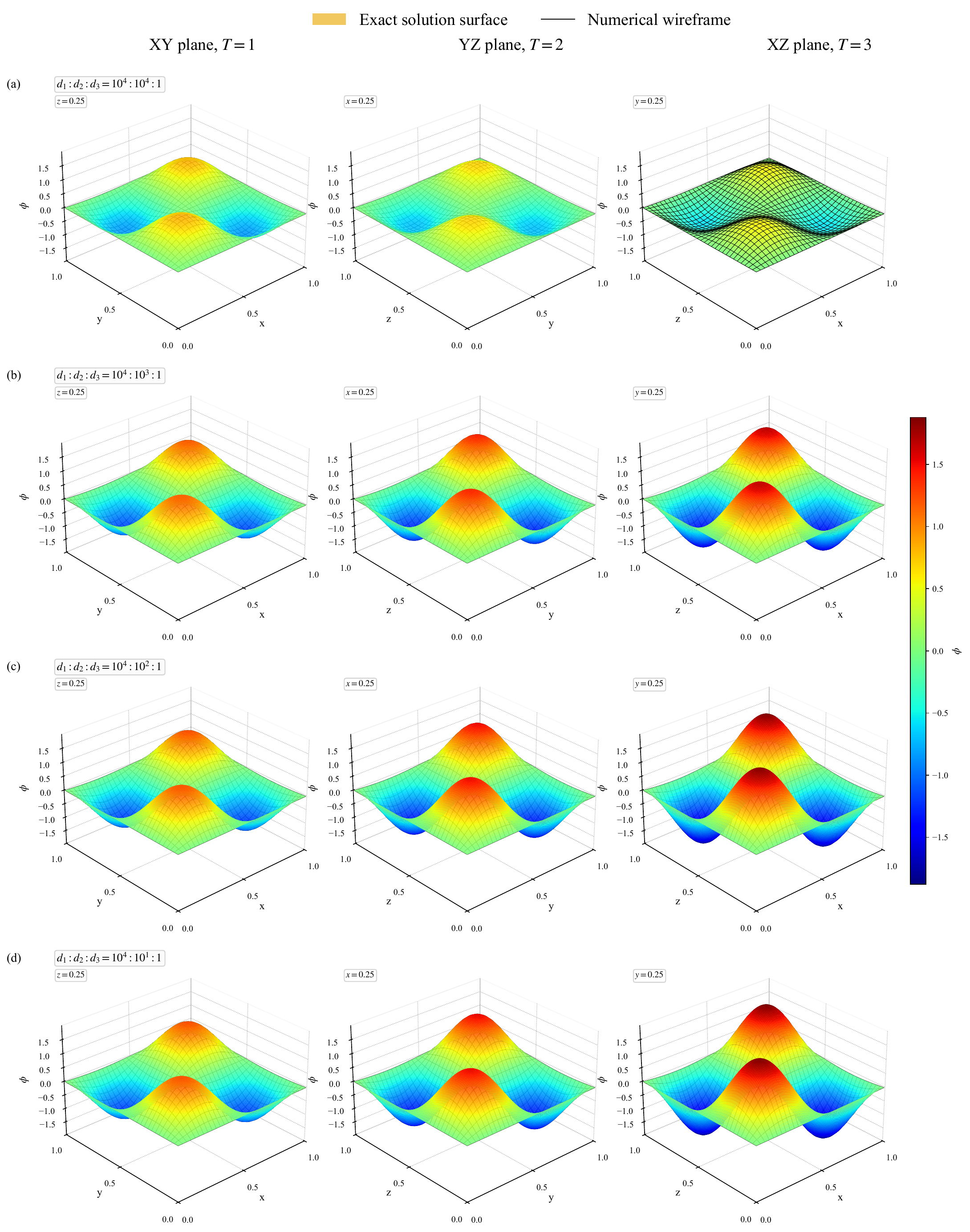}
    \caption{Numerical and analytical solutions for the three-dimensional anisotropic advection--diffusion equation with constant velocity and variable diffusion tensor at $Pe=10^6$. Rows (a)--(d) correspond to the four baseline diffusivity ratios $d_1:d_2:d_3=10^4\!:\!10^4\!:\!1$, $10^4\!:\!10^3\!:\!1$, $10^4\!:\!10^2\!:\!1$, and $10^4\!:\!10^1\!:\!1$. Columns show the $xy$ section at $z=0.25$ and $t=1$, the $yz$ section at $x=0.25$ and $t=2$, and the $xz$ section at $y=0.25$ and $t=3$. Colored surfaces denote the exact solution and black wireframes denote the numerical solution.}
    \label{fig:chai3d_variable_diffusion}
\end{figure}

We performed simulations on a fixed $128^3$ lattice and present the results for the high-P\'eclet case $Pe=10^6$, where
\begin{equation}
Pe=\frac{L|u_x^{\mathrm{lat}}|}{\kappa_{\mathrm{base}}}=\frac{L|u_x^{\mathrm{lat}}|}{d_3},\qquad
L=128,\qquad
\kappa_{\mathrm{base}}=d_3=10^{-6}.
\end{equation}
The imposed physical velocity is $\mathbf{u}=(1,1,1)$. In the lattice-fixed runs, this gives the lattice velocity $\mathbf{u}^{\mathrm{lat}}=(0.0078125,\,0.0078125,\,0.0078125)$. The largest and smallest baseline diffusivities are fixed at $d_1=10^{-2}$ and $d_3=\kappa_{\mathrm{base}}=10^{-6}$. The intermediate diffusivity is varied over
\begin{equation}
(d_1,d_2,d_3)\in\left\{
(10^{-2},10^{-2},10^{-6}),\,
(10^{-2},10^{-3},10^{-6}),\,
(10^{-2},10^{-4},10^{-6}),\,
(10^{-2},10^{-5},10^{-6})
\right\}.
\end{equation}
To maintain positivity of the transported populations in this source-driven problem, the computation is performed for a shifted scalar field $\psi=\phi+s$, where the constant $s>0$ is chosen so that $\psi$ remains strictly positive throughout the simulation. Because the imposed velocity is constant and divergence-free, and because the diffusion operator acts on gradients, the shifted variable satisfies
\begin{equation}
\partial_t \psi+\nabla\cdot(\psi\mathbf{u})=\nabla\cdot(\mathbf{K}\nabla\psi)+R,
\end{equation}
with the same source term $R$ as the original equation for $\phi$. The numerical results reported here are obtained with a symmetric second-order splitting in which the source is evaluated at the midpoint of each full time step,
\begin{equation}
\mathcal{S}_{n+1/2}\!\left(\frac{\Delta t}{2}\right)\rightarrow
\mathcal{T}(\Delta t)\rightarrow
\mathcal{S}_{n+1/2}\!\left(\frac{\Delta t}{2}\right),
\end{equation}
where $\mathcal{T}$ denotes the source-free anisotropic ADE-KBC transport carrier update and $\mathcal{S}_{n+1/2}$ denotes a source half-step with the source frozen at the same midpoint time $t_{n+1/2}=t_n+\Delta t/2$ on both sides of the carrier step. Thus the two half-steps approximate the source integral over the full step by the midpoint rule,
\[
\int_{t_n}^{t_{n+1}}R(\mathbf{x},t)\,dt
=\Delta t\,R(\mathbf{x},t_{n+1/2})+O(\Delta t^3),
\]
which gives the second-order temporal accuracy of the split source update~\citep{Strang1968SJNA}. Following lattice Boltzmann source-term treatments for advection--diffusion solvers~\citep{HamilaJemniPerre2023IJP}, each half-step incorporates the source contribution directly in population space as
\begin{equation}
g_i^{\,\star}=g_i + \frac{\Delta t}{2}R(\mathbf{x},t_{n+1/2})\,
\mathcal{E}_i(\mathbf{u}^{\mathrm{lat}}),
\end{equation}
where
\begin{equation}
\mathcal{E}_i(\mathbf{u}^{\mathrm{lat}})=
W_i\left(
1+\frac{\mathbf{c}_i\cdot\mathbf{u}^{\mathrm{lat}}}{c_s^2}
+\frac{(\mathbf{c}_i\cdot\mathbf{u}^{\mathrm{lat}})^2}{2c_s^4}
-\frac{\mathbf{u}^{\mathrm{lat}}\cdot\mathbf{u}^{\mathrm{lat}}}{2c_s^2}
\right),
\end{equation}
is the second-order equilibrium polynomial evaluated at unit density. This construction distributes the source over the discrete velocity set in a form consistent with the transport equilibrium and preserves second-order temporal accuracy of the split update.
Figure~\ref{fig:chai3d_variable_diffusion} compares the analytical and numerical solutions on three representative sections of the cube: the $xy$ plane at $z=0.25$ and $T=1$, the $yz$ plane at $x=0.25$ and $T=2$, and the $xz$ plane at $y=0.25$ and $T=3$. In each panel, the exact solution is shown as a colored surface and the numerical result is superposed as a wireframe.

As seen from Fig.~\ref{fig:chai3d_variable_diffusion}, the numerical results are very close to the analytical solutions for all four anisotropy ratios and at all three output times. For $d_1:d_2:d_3=10^4\!:\!10^4\!:\!1$, the solution amplitude decays in time because the growth rate $A$ is negative. For the other three cases, the extrema increase gradually as $d_2$ decreases and $A$ becomes positive. The overlap between the wireframes and the analytical surfaces remains uniformly tight near the extrema and the zero level sets. The same level of agreement appears on the $xy$, $xz$, and $yz$ planes. On the $xy$ and $xz$ sections, the spatial variation of $K_{xx}$ enters directly. The agreement on the $yz$ plane confirms that the present formulation captures the coupled three-dimensional effect of advection, anisotropic diffusion, and source forcing.

To quantitatively measure the deviations between numerical and analytical solutions, we computed the global relative errors (GREs) at the final time $t=3$. For the four diffusivity ratios $10^4\!:\!10^4\!:\!1$, $10^4\!:\!10^3\!:\!1$, $10^4\!:\!10^2\!:\!1$, and $10^4\!:\!10^1\!:\!1$, the GREs are $6.220\times10^{-4}$, $6.161\times10^{-4}$, $6.155\times10^{-4}$, and $6.154\times10^{-4}$, respectively. The corresponding $L_2$ errors lie between $1.26\times10^{-4}$ and $4.09\times10^{-4}$. The $L_\infty$ errors remain below $1.31\times10^{-3}$. In addition, the absolute mass drift of $\phi$ at $t=3$ is below $2.1\times10^{-11}$ in every case. These error levels are sufficiently small to demonstrate that the present ELBM formulation is accurate for three-dimensional anisotropic advection--diffusion with a variable diffusion tensor and a source term.

\FloatBarrier

%% file: sections/application.tex
\section{Application}
\label{sec:application}

\subsection{Taylor dispersion of elongated rods in plane Poiseuille flow}
\label{sec:application-rods}

We consider the Taylor dispersion of elongated Brownian rods in a plane Poiseuille flow using the continuum model and asymptotic reduction reported by Kumar \textit{et al.}~\citep{KumarEtAl2021}. The problem generates a wall-normal drift and a position-dependent anisotropic diffusion tensor through the local orientational statistics of the rods. It therefore provides a stringent test for the present ELBM in a setting where the transport coefficients follow directly from the microscale orientational dynamics of the rods.

\begin{figure}[pos=H]
    \centering
    \includegraphics[width=0.98\textwidth]{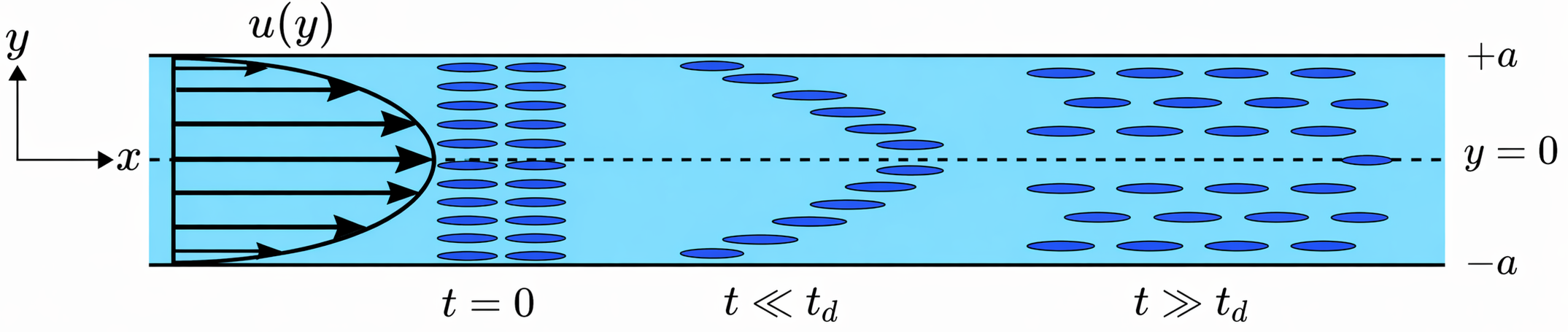}
    \caption{Schematic of Taylor dispersion of elongated rods in a plane Poiseuille flow. A Gaussian plug is initially localized in the streamwise direction and uniformly distributed across the channel width. Shear-induced rotation and anisotropic diffusion redistribute the rods across the shear layers. At long times, this coupling yields an effective axial transport with modified mean speed and enhanced longitudinal dispersion.}
    \label{fig:rods_schematic}
\end{figure}

\begin{figure}[pos=H]
    \centering
    \includegraphics[width=0.98\textwidth]{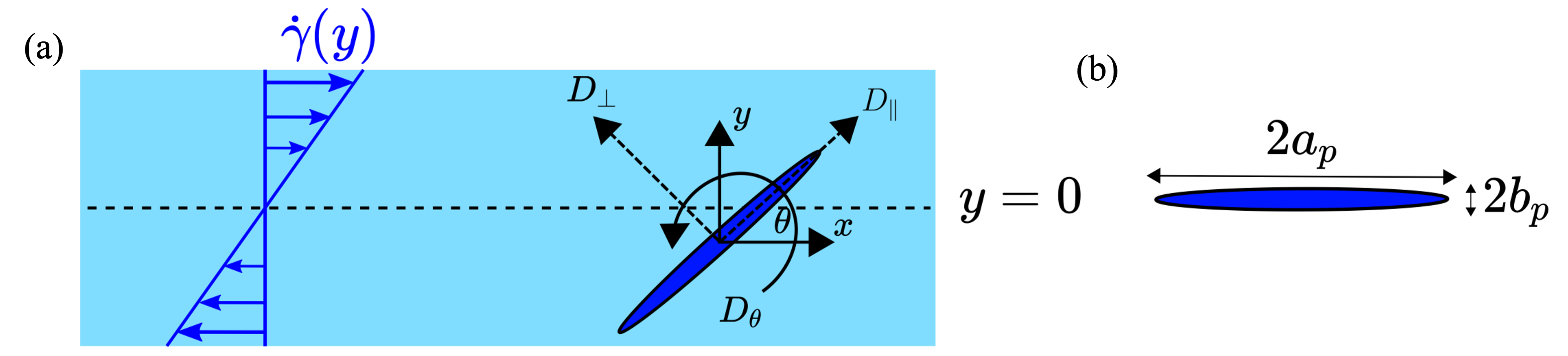}
    \caption{Geometric and transport quantities for an elongated Brownian rod in a plane Poiseuille flow~\citep{KumarEtAl2021}. The local shear rate $\dot{\gamma}(y)$ drives Jeffery rotation, $\theta$ is the in-plane orientation angle, $D_{\parallel}$ and $D_{\perp}$ are the translational diffusivities parallel and perpendicular to the rod axis, and $D_{\theta}$ is the rotational diffusivity. The semimajor and semiminor axes are denoted by $a_p$ and $b_p$, so the aspect ratio is $p=a_p/b_p$.}
    \label{fig:rods_definitions}
\end{figure}

The geometric and transport quantities entering the model are summarized in Fig.~\ref{fig:rods_definitions}. We consider a channel of half-width $a$ with the dimensional Poiseuille profile
\begin{equation}
u_f(y)=U\left(1-\frac{y^2}{a^2}\right), \qquad -a \le y \le a,
\label{eq:app-poiseuille}
\end{equation}
where $U$ is the centerline velocity. A rod with aspect ratio $p=a_p/b_p$, defined from the semimajor axis $a_p$ and semiminor axis $b_p$ in Fig.~\ref{fig:rods_definitions}, is described by the in-plane orientation angle $\theta$, measured with respect to the streamwise direction. Lengths are nondimensionalized by $a$, time by the transverse diffusive time $t_d=a^2/\bar D$, and the flow by $U$, where
\begin{equation}
\bar D=\frac{D_{\parallel}+D_{\perp}}{2}
\end{equation}
is the orientation-averaged translational diffusivity. Here $D_{\parallel}$ and $D_{\perp}$ denote the translational diffusion coefficients parallel and perpendicular to the rod axis shown in Fig.~\ref{fig:rods_definitions}, and $D_{\theta}$ denotes the rotational diffusivity. For a prolate spheroid, the translational diffusivities follow Perrin's formulas as summarized by Happel and Brenner~\citep{Perrin1936,HappelBrenner2012}, while the rotational diffusivity follows Perrin's rotational expression with the correction reported by Koenig~\citep{Perrin1934,Koenig1975}:
\begin{subequations}
\label{eq:app-perrin}
\begin{align}
D_{\parallel}
&=
\frac{k_B T}{16\pi\mu a_p}\,
p\left[
-\frac{2p}{p^2-1}
+\frac{2p^2-1}{(p^2-1)^{3/2}}
\log\!\left(\frac{p+\sqrt{p^2-1}}{p-\sqrt{p^2-1}}\right)
\right],
\label{eq:app-perrin-par}
\\
D_{\perp}
&=
\frac{k_B T}{16\pi\mu a_p}\,
p\left[
\frac{p}{p^2-1}
+\frac{2p^2-3}{(p^2-1)^{3/2}}
\log\!\left(p+\sqrt{p^2-1}\right)
\right],
\label{eq:app-perrin-perp}
\\
D_{\theta}
&=
\frac{3k_B T}{16\pi\mu a_p^3}\,
\frac{p^4}{p^4-1}
\left[
\frac{2p^2-1}{p\sqrt{p^2-1}}
\log\!\left(p+\sqrt{p^2-1}\right)-1
\right].
\label{eq:app-perrin-rot}
\end{align}
\end{subequations}
Here $k_B$ is Boltzmann's constant, $T$ is the absolute temperature, and $\mu$ is the dynamic viscosity. In the present nondimensionalization, $D_{\parallel}/\bar D$ and $D_{\perp}/\bar D$ determine the anisotropic translational tensor in Eqs.~\eqref{eq:app-diffusion-tensor}--\eqref{eq:app-diffusion-components}, while $D_{\theta}$ enters through the rotational P\'eclet number. The relevant control parameters are the translational and rotational P\'eclet numbers
\begin{equation}
Pe=\frac{Ua}{\bar D}, \qquad
Pe_r=\frac{U}{aD_{\theta}},
\label{eq:app-pe-per}
\end{equation}
with $D_{\theta}$ the rotational diffusivity. In this scaling the dimensionless streamwise velocity is
\begin{equation}
u(y)=1-y^2,
\end{equation}
and the Jeffery rotation rate for a prolate spheroid confined to one rotational degree of freedom is
\begin{equation}
\omega(y,\theta)=
-2y\,\frac{p^2\sin^2\theta+\cos^2\theta}{p^2+1}.
\label{eq:app-omega}
\end{equation}

The translational diffusion tensor of a rod depends on its instantaneous orientation,
\begin{equation}
\mathbf{D}(\theta)=D_{\perp}\mathbf{I}+\left(D_{\parallel}-D_{\perp}\right)\mathbf{e}\mathbf{e}^{\mathsf T},
\qquad
\mathbf{e}=(\cos\theta,\sin\theta)^{\mathsf T},
\label{eq:app-diffusion-tensor}
\end{equation}
which gives the Cartesian components
\begin{equation}
\mathbf{D}(\theta)=
\begin{bmatrix}
D_{\parallel}\cos^2\theta+D_{\perp}\sin^2\theta &
\left(D_{\parallel}-D_{\perp}\right)\sin\theta\cos\theta
\\[4pt]
\left(D_{\parallel}-D_{\perp}\right)\sin\theta\cos\theta &
D_{\parallel}\sin^2\theta+D_{\perp}\cos^2\theta
\end{bmatrix}.
\label{eq:app-diffusion-components}
\end{equation}
For each aspect ratio $p$, these expressions fix the anisotropic tensor entering the transport model and the rotational time scale that appears in $Pe_r$.

The reference curves are obtained from the long-time asymptotic theory for the regime $Pe_r/Pe \ll 1$, where the orientational dynamics equilibrate more rapidly than the axial spreading. Introducing the moving coordinate $X=x-u_m t$ and the slow variables $\xi=\varepsilon^2 X$ and $T=\varepsilon^2 t$ with $\varepsilon=Pe_r/Pe$, the particle probability density is written as
\begin{equation}
P(x,y,\theta,t)=\frac{1}{N}\,g(\theta;y)\,\mathcal{C}(\xi,y,T),
\label{eq:app-separation}
\end{equation}
where $N$ is the total number of particles, $g(\theta;y)$ is the steady orientational distribution at a fixed shear layer, and $\mathcal{C}$ is the slowly varying concentration field. The leading-order orientational problem is
\begin{equation}
\frac{\partial^2 g}{\partial \theta^2}
-Pe_r\frac{\partial}{\partial\theta}\!\left[\omega(y,\theta)g\right]=0,
\qquad
\int_{0}^{\pi} g(\theta;y)\,d\theta = 1.
\label{eq:app-g}
\end{equation}
with $\pi$-periodicity in $\theta$. In practice, Eq.~\eqref{eq:app-g} is solved numerically through a truncated Fourier representation on $\theta\in[0,\pi)$,
\begin{equation}
g(\theta;y)=\frac{1}{\pi}
+\sum_{n=1}^{M}\left[a_n(y)\cos(2n\theta)+b_n(y)\sin(2n\theta)\right],
\label{eq:app-fourier}
\end{equation}
where $M$ is the truncation order. The coefficients are obtained by a Fourier-collocation solve.

Once $g(\theta;y)$ is known, the effective transport coefficients follow from orientational averaging. Writing
\begin{equation}
\langle f\rangle = \int_0^{\pi} f(\theta)\,d\theta,
\qquad
\overline{q}=\frac{1}{2}\int_{-1}^{1} q(y)\,dy,
\end{equation}
the effective lateral diffusivity and migration velocity are
\begin{equation}
D_y(y)=\left\langle D_{yy}(\theta)\,g(\theta;y)\right\rangle,
\qquad
v_d(y)=\frac{dD_y}{dy}.
\label{eq:app-dy-vd}
\end{equation}
The leading-order concentration profile satisfies $\mathcal{C}^{(0)}(\xi,y,T)=C_m(\xi,T)/D_y(y)$, and the corresponding mean transport speed is
\begin{equation}
u_m^{(0)}=\frac{\overline{D_y^{-1}u}}{\overline{D_y^{-1}}}.
\label{eq:app-um0}
\end{equation}
Defining
\begin{equation}
G(y)=\int_{-1}^{y}\int_{-1}^{z}
D_y^{-1}(y')\left[u(y')-u_m^{(0)}\right]\,dy'\,dz,
\label{eq:app-G}
\end{equation}
the effective dispersion coefficient is
\begin{equation}
\kappa=
-\frac{\overline{D_y^{-1}G\left(u-u_m^{(0)}\right)}}{\overline{D_y^{-1}}},
\label{eq:app-kappa}
\end{equation}
and the $O(Pe^{-1})$ correction to the mean speed yields
\begin{equation}
u_m=
\frac{\overline{D_y^{-1}u}}{\overline{D_y^{-1}}}
-\frac{1}{2Pe\,\overline{D_y^{-1}}}
\left[
D_y^{-1}\left\langle gD_{xy}\right\rangle
\right]_{y=-1}^{y=1}.
\label{eq:app-um}
\end{equation}
The long-time, laterally averaged concentration therefore obeys the one-dimensional transport equation
\begin{equation}
\frac{\partial C_m}{\partial t}
+Pe\,u_m\frac{\partial C_m}{\partial x}
=\kappa\,Pe^2\frac{\partial^2 C_m}{\partial x^2}.
\label{eq:app-effective-transport}
\end{equation}
This procedure yields semi-analytic predictions for $u_m$ and $\kappa$: the closure formulas are explicit, while the orientational boundary-value problem is solved numerically.

The ELBM validation is carried out by solving the corresponding two-dimensional advection--diffusion equation with wall-normal drift and a spatially varying diffusion tensor,
\begin{equation}
\partial_t C
+\nabla\cdot\!\left(\mathbf{u}_{\mathrm{adv}}(y)C\right)
=
\nabla\cdot\!\left(\mathbf{K}(y)\nabla C\right),
\label{eq:app-elbm-pde}
\end{equation}
where
\begin{equation}
\mathbf{u}_{\mathrm{adv}}(y)=
\begin{bmatrix}
U\,u(y) \\
-\dfrac{\bar D}{a}\,v_d(y)
\end{bmatrix},
\qquad
\mathbf{K}(y)=
\bar D
\begin{bmatrix}
D_{xx}(y) & D_{xy}(y)\\
D_{xy}(y) & D_y(y)
\end{bmatrix}.
\label{eq:app-elbm-fields}
\end{equation}
Here $D_{xx}(y)=\langle D_{xx}(\theta)\,g(\theta;y)\rangle$ and $D_{xy}(y)=\langle D_{xy}(\theta)\,g(\theta;y)\rangle$ are the orientationally averaged streamwise and cross components of the diffusivity tensor. The wall-normal component is $D_{yy}(y)=\langle D_{yy}(\theta)\,g(\theta;y)\rangle$, which was introduced above in the abbreviated form $D_y(y)$ because it enters directly in the migration velocity $v_d=dD_y/dy$ and in the long-time closure formulas.
The vector $\mathbf{u}_{\mathrm{adv}}$ collects the two first-order transport contributions in the reduced model. Its streamwise component $Uu(y)$ is the imposed Poiseuille advection. Its wall-normal component $-(\bar D/a)\,v_d(y)$ represents the migration drift associated with the spatial variation of the effective lateral diffusivity $D_y(y)$. This drift appears when the orientationally averaged wall-normal transport is written in conservative form: the $y$-dependence of $D_y$ produces a term proportional to $v_d(y)C$, which enters the governing equation at the same differential order as advection. Writing these two contributions as a vector provides a compact statement of the full first-order transport in the $(x,y)$ plane.
The axial direction is periodic and the walls at $y=\pm a$ satisfy the anisotropic no-flux condition. The initial condition is a Gaussian plug that is uniform across the channel width,
\begin{equation}
C(x,y,0)=
\exp\!\left[-\frac{(x-x_0)^2}{2\sigma_x^2}\right].
\label{eq:app-initial}
\end{equation}
In Eq.~\eqref{eq:app-initial}, $x_0$ is the initial streamwise center of the plug and $\sigma_x$ is its initial standard deviation. The ELBM calculations use the D2Q9 lattice together with the entropic anisotropic formulation introduced in Sec.~\ref{sec:method-collision}. The computational domain is $N_x\times N_y=30000\times33$, the imposed flow P\'eclet number is $Pe=1000$, the lattice centerline speed is $U_{\mathrm{lbm}}=0.02$, and the initial Gaussian width is $\sigma_x=24$ lattice units. With $a_{\mathrm{lbm}}=N_y/2=16.5$, these choices give $\bar D_{\mathrm{lbm}}=U_{\mathrm{lbm}}a_{\mathrm{lbm}}/Pe=3.3\times10^{-4}$ and $t_{\mathrm{cross}}=a_{\mathrm{lbm}}^2/\bar D_{\mathrm{lbm}}=8.25\times10^{5}$. We consider aspect ratios $p=2$, 10, 100, 1000, and $\infty$, and rotational P\'eclet numbers $Pe_r=10^{-2}$, $10^{-1}$, 1, 10, $10^2$, and $10^3$.
Following Kumar et al.~\citep{KumarEtAl2021}, the physically relevant rod-dispersion regime satisfies $Pe_r\ll Pe$ (and, more generally, $Pe_r<Pe$). The endpoint $Pe_r=10^3$ at $Pe=1000$ is therefore retained as a numerical robustness and coverage test of the anisotropic solver rather than as a physically representative parameter point.
Each run is advanced for $1.2\times10^6$ steps, corresponding to $t/t_{\mathrm{cross}}=1.455$, and the long-time slopes of $\langle x\rangle$ and $\sigma_x^2$ are fitted over the interval $t/t_{\mathrm{cross}}\ge 0.727$. The reported dispersion is normalized by the classical Taylor--Aris value for spherical tracers,
\begin{equation}
\kappa_s=\frac{8}{945}.
\label{eq:app-kappas}
\end{equation}

\begin{figure}[pos=H]
    \centering
    \includegraphics[width=0.98\textwidth]{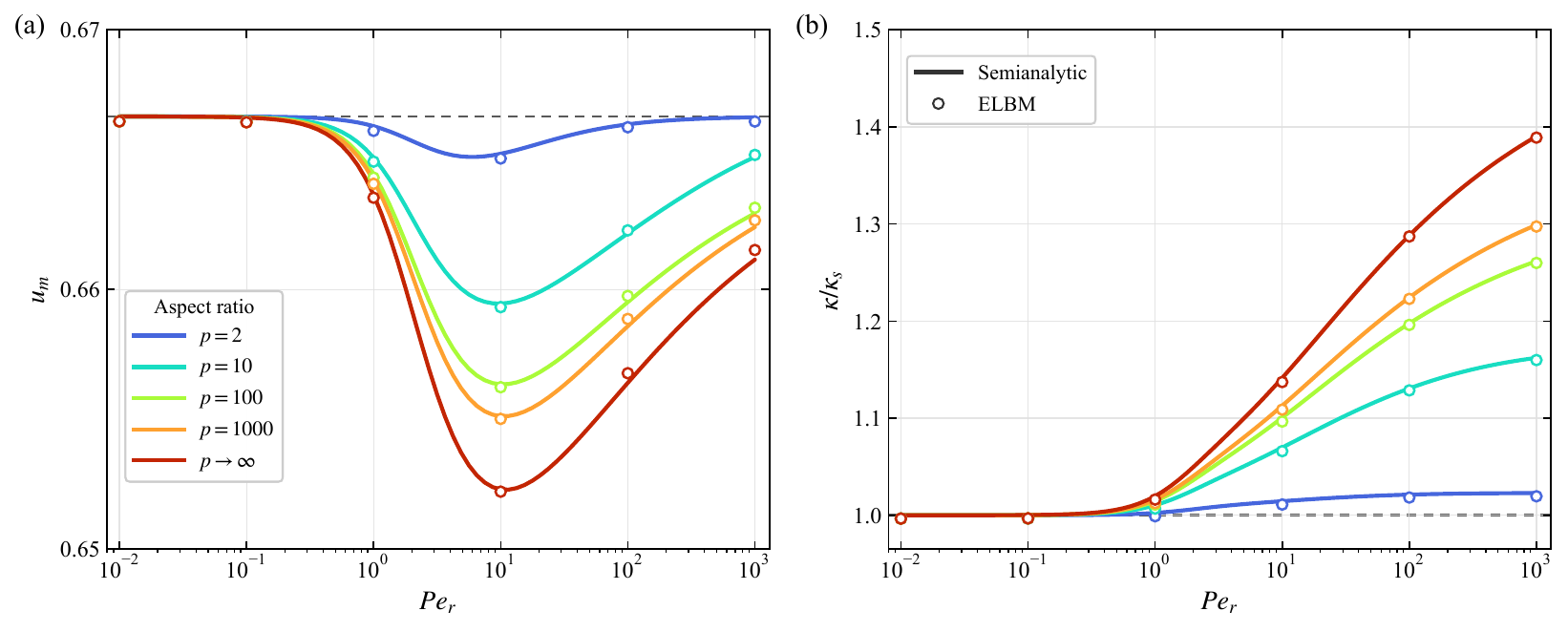}
    \caption{Comparison between the semi-analytic closure and ELBM for Taylor dispersion of elongated rods at $Pe=1000$. Panel (a) shows the dimensionless mean particle speed $u_m$ as a function of $Pe_r$; panel (b) shows the normalized effective dispersion $\kappa/\kappa_s$. Solid lines denote the semi-analytic predictions obtained from Eqs.~\eqref{eq:app-g}--\eqref{eq:app-effective-transport}, and open circles are ELBM measurements extracted from the long-time growth of the first two axial moments. The dashed horizontal lines indicate the spherical-reference limits $u_m=2/3$ and $\kappa/\kappa_s=1$.}
    \label{fig:rods_results}
\end{figure}

The trends in Fig.~\ref{fig:rods_results} are consistent with the physical picture in Fig.~\ref{fig:rods_schematic}. When $Pe_r\ll1$, rotational diffusion dominates Jeffery rotation, the rods sample orientations nearly isotropically, and the transport approaches the spherical limit: all curves begin close to $u_m=2/3$ and $\kappa/\kappa_s=1$. As $Pe_r$ increases, the rods spend more time aligned with the streamwise direction. This reduces their ability to diffuse across the shear layers, strengthens the wall-normal migration toward low-mobility regions near the walls, and lowers the mean axial speed. The minimum in $u_m$ occurs around $Pe_r\approx10$ and becomes deeper as the aspect ratio increases. In the slender-body limit, the semi-analytic prediction reaches $u_m\approx0.652$ at $Pe_r=10$, while the ELBM point remains on the same curve within plotting accuracy.

The dispersion enhancement in Fig.~\ref{fig:rods_results}(b) grows monotonically with both $Pe_r$ and $p$. At $Pe_r=1000$, the semi-analytic values of $\kappa/\kappa_s$ are approximately $1.023$, $1.162$, $1.262$, $1.299$, and $1.390$ for $p=2$, $10$, $100$, $1000$, and $\infty$, respectively. The ELBM reproduces these values closely over the tested parameter range. The largest relative discrepancy is $5.7\times10^{-4}$ for $u_m$ and $3.7\times10^{-3}$ for $\kappa/\kappa_s$. The agreement in both panels shows that the present ELBM recovers the long-time transport generated by orientation-dependent anisotropic diffusion, cross-diffusion, and shear-induced migration in this Brownian-rod dispersion problem.

\subsection{Anisotropic thermal-conductivity measurement}
\label{sec:application-thermal-conductivity}

\subsubsection{Steady heat conduction in a rotated anisotropic solid}
\label{sec:application-thermal-conductivity-steady}

We consider a homogeneous solid whose principal thermal axes are inclined with respect to the Cartesian grid. Anisotropic thermal-conductivity measurements are needed in oriented polymers and heterogeneous geomaterials, where the apparent conductivity depends on material orientation and, in some cases, measurement scale~\citep{Kurabayashi2001Thermophys,LiChenChen2024BEE}. The problem below is the thermal-conductivity counterpart of a guarded-plate measurement: a temperature difference is imposed between two planes normal to the $x$ direction, and the material tensor is rotated so that the imposed gradient couples to both diagonal and off-diagonal conductivity entries. The steady heat equation and Fourier law are
\begin{equation}
\nabla\cdot\mathbf{q}=0,
\qquad
\mathbf{q}=-\mathbf{K}\nabla T,
\label{eq:app-heat-fourier}
\end{equation}
where $\mathbf{K}$ is given in $\mathrm{W\,m^{-1}K^{-1}}$. The domain is a $128^3$ cube. The planes $x=0$ and $x=L_x$ are held at temperatures $T_h$ and $T_c$, while the $y$ and $z$ directions are periodic to remove side-wall effects. The calculation therefore contains a single imposed macroscopic gradient,
\begin{equation}
G_x=\frac{\Delta T}{L_x}.
\label{eq:app-heat-gradient}
\end{equation}
Here $\Delta T=T_h-T_c$. The computation is nondimensionalized by $\Delta T$, and the dimensional heat flux is restored with the physical conductivities used below. In Table~\ref{tab:thermal-conductivity-num}, the reported fluxes correspond to $G_x=1/128~\mathrm{K\,m^{-1}}$; a different experimental gradient would simply multiply every heat-flux entry by $G_x/(1/128)$, leaving the recovered conductivity entries unchanged. This makes the test a direct check of the tensor entries, expressed in Cartesian coordinates, that multiply the imposed $x$-direction gradient.

The conductivity tensor is obtained by rotating the principal tensor,
\begin{equation}
\mathbf{K}
=
\mathbf{R}_z(\alpha_E)\mathbf{R}_y(\beta_E)\mathbf{R}_z(\gamma_E)
\begin{bmatrix}
k_1&0&0\\
0&k_2&0\\
0&0&k_3
\end{bmatrix}
\mathbf{R}_z^{\mathsf T}(\gamma_E)\mathbf{R}_y^{\mathsf T}(\beta_E)\mathbf{R}_z^{\mathsf T}(\alpha_E),
\label{eq:app-heat-rotated-k}
\end{equation}
using the same ZYZ Euler-angle convention as in Sec.~\ref{sec:numerical-gaussian3d}. The largest and smallest principal conductivities are fixed at
\begin{equation}
k_1=100~\mathrm{W\,m^{-1}K^{-1}},
\qquad
k_3=0.01~\mathrm{W\,m^{-1}K^{-1}},
\end{equation}
and the intermediate value is set to $k_2=100$, 10, 1, and $0.1~\mathrm{W\,m^{-1}K^{-1}}$. These values give the ratios $10^4:10^4:1$, $10^4:10^3:1$, $10^4:10^2:1$, and $10^4:10^1:1$. The Euler angles are changed from row to row so that the comparison samples a broad range of $K_{xx}$, positive and negative off-diagonal coupling, and different balances between $K_{xy}$ and $K_{xz}$.

For a uniform tensor and periodic transverse boundaries, the exact steady solution is one-dimensional,
\begin{equation}
T(x)=T_h-G_x x,
\qquad
\nabla T=\left(-G_x,0,0\right)^{\mathsf T}.
\end{equation}
Fourier's law then gives the three flux contributions driven by this $x$-gradient,
\begin{equation}
\begin{aligned}
q_{xx}^{\mathrm{ex}}&=K_{xx}G_x,\\
q_{xy}^{\mathrm{ex}}&=K_{xy}G_x,\\
q_{xz}^{\mathrm{ex}}&=K_{xz}G_x.
\end{aligned}
\label{eq:app-heat-exact-q}
\end{equation}
The notation $q_{xx}$, $q_{xy}$, and $q_{xz}$ keeps the heat-flux components tied to the conductivity entries multiplying $G_x$; no separate $y$- or $z$-direction temperature gradient is applied. In the ELBM data, the corresponding numerical heat-flux components are evaluated as
\begin{equation}
\begin{aligned}
q_{xx}^{\mathrm{num}}\!\left(x_{j+1/2}\right)&=K_{xx}\frac{T_j-T_{j+1}}{\Delta x},\\
q_{xy}^{\mathrm{num}}\!\left(x_{j+1/2}\right)&=K_{xy}\frac{T_j-T_{j+1}}{\Delta x},\\
q_{xz}^{\mathrm{num}}\!\left(x_{j+1/2}\right)&=K_{xz}\frac{T_j-T_{j+1}}{\Delta x}.
\end{aligned}
\label{eq:app-heat-num-flux}
\end{equation}
At steady state, these numerical fluxes are constant over the interior of the sample. The conductivity entries are recovered from the corresponding interior values as
\begin{equation}
K_{x\ell}^{\mathrm{num}}
=
\frac{\bar q_{x\ell}^{\mathrm{num}}}{G_x},
\qquad \ell\in\{x,y,z\}.
\label{eq:app-heat-recovered-k}
\end{equation}

\begin{table}[pos=H]
    \centering
    \rmfamily
    \caption{Numerical recovery of the heat-flux and thermal-conductivity entries coupled to the imposed $x$-gradient. Conductivities are in $\mathrm{W\,m^{-1}K^{-1}}$, and heat fluxes are in $\mathrm{W\,m^{-2}}$ for $G_x=1/128~\mathrm{K\,m^{-1}}$. Each heat-flux and conductivity cell gives the exact value above the numerical value.}
    \label{tab:thermal-conductivity-num}
    \begingroup
    \small
    \setlength{\tabcolsep}{2.6pt}
    \renewcommand{\arraystretch}{1.08}
    \begin{tabular*}{\textwidth}{@{\extracolsep{\fill}}lccccccc@{}}
        \toprule
        $k_1:k_2:k_3$ &
        $(\alpha_E,\beta_E,\gamma_E)$ &
        \makecell[c]{$q_{xx}$\\ ex./num.} &
        \makecell[c]{$q_{xy}$\\ ex./num.} &
        \makecell[c]{$q_{xz}$\\ ex./num.} &
        \makecell[c]{$K_{xx}$\\ ex./num.} &
        \makecell[c]{$K_{xy}$\\ ex./num.} &
        \makecell[c]{$K_{xz}$\\ ex./num.} \\
        \midrule
        $10^4:10^4:1$ & $(120^\circ,45^\circ,60^\circ)$
        & \makecell[c]{$0.68360$\\$0.68350$} & \makecell[c]{$0.16913$\\$0.16910$} & \makecell[c]{$0.19529$\\$0.19526$}
        & \makecell[c]{$87.501$\\$87.488$} & \makecell[c]{$21.648$\\$21.645$} & \makecell[c]{$24.997$\\$24.994$} \\
        $10^4:10^3:1$ & $(135^\circ,35^\circ,75^\circ)$
        & \makecell[c]{$0.55309$\\$0.55302$} & \makecell[c]{$0.32505$\\$0.32500$} & \makecell[c]{$0.11287$\\$0.11286$}
        & \makecell[c]{$70.796$\\$70.786$} & \makecell[c]{$41.606$\\$41.600$} & \makecell[c]{$14.448$\\$14.445$} \\
        $10^4:10^2:1$ & $(105^\circ,50^\circ,25^\circ)$
        & \makecell[c]{$0.24919$\\$0.24916$} & \makecell[c]{$-0.19486$\\$-0.19484$} & \makecell[c]{$0.30115$\\$0.30111$}
        & \makecell[c]{$31.896$\\$31.892$} & \makecell[c]{$-24.942$\\$-24.939$} & \makecell[c]{$38.548$\\$38.542$} \\
        $10^4:10^1:1$ & $(145^\circ,55^\circ,55^\circ)$
        & \makecell[c]{$0.42709$\\$0.42703$} & \makecell[c]{$0.27853$\\$0.27850$} & \makecell[c]{$0.27139$\\$0.27135$}
        & \makecell[c]{$54.667$\\$54.660$} & \makecell[c]{$35.652$\\$35.647$} & \makecell[c]{$34.737$\\$34.733$} \\
        \bottomrule
    \end{tabular*}
    \endgroup
\end{table}

The numerical values in Table~\ref{tab:thermal-conductivity-num} recover the prescribed conductivity entries with relative discrepancies of $1.46\times10^{-4}$, $1.41\times10^{-4}$, $1.40\times10^{-4}$, and $1.21\times10^{-4}$ in the four cases. The dimensional fluxes are consistent with the imposed gradient; for example, the first row gives $q_{xx}^{\mathrm{num}}=0.68350~\mathrm{W\,m^{-2}}$, which corresponds to $K_{xx}^{\mathrm{num}}=87.488~\mathrm{W\,m^{-1}K^{-1}}$ after division by $1/128~\mathrm{K\,m^{-1}}$. The third row also checks the sign of the off-diagonal term: the selected rotation gives $K_{xy}<0$, and the recovered $q_{xy}^{\mathrm{num}}$ carries the same sign under the imposed positive $G_x$. These results show that the method recovers $K_{xx}$, $K_{xy}$, and $K_{xz}$ from the physical heat flux generated by a single $x$-direction temperature gradient.

\subsubsection{Effective heat conduction in anisotropic porous cube-array materials}
\label{sec:application-thermal-conductivity-cubes}

\begin{figure}[pos=H]
    \centering
    \includegraphics[width=0.98\textwidth]{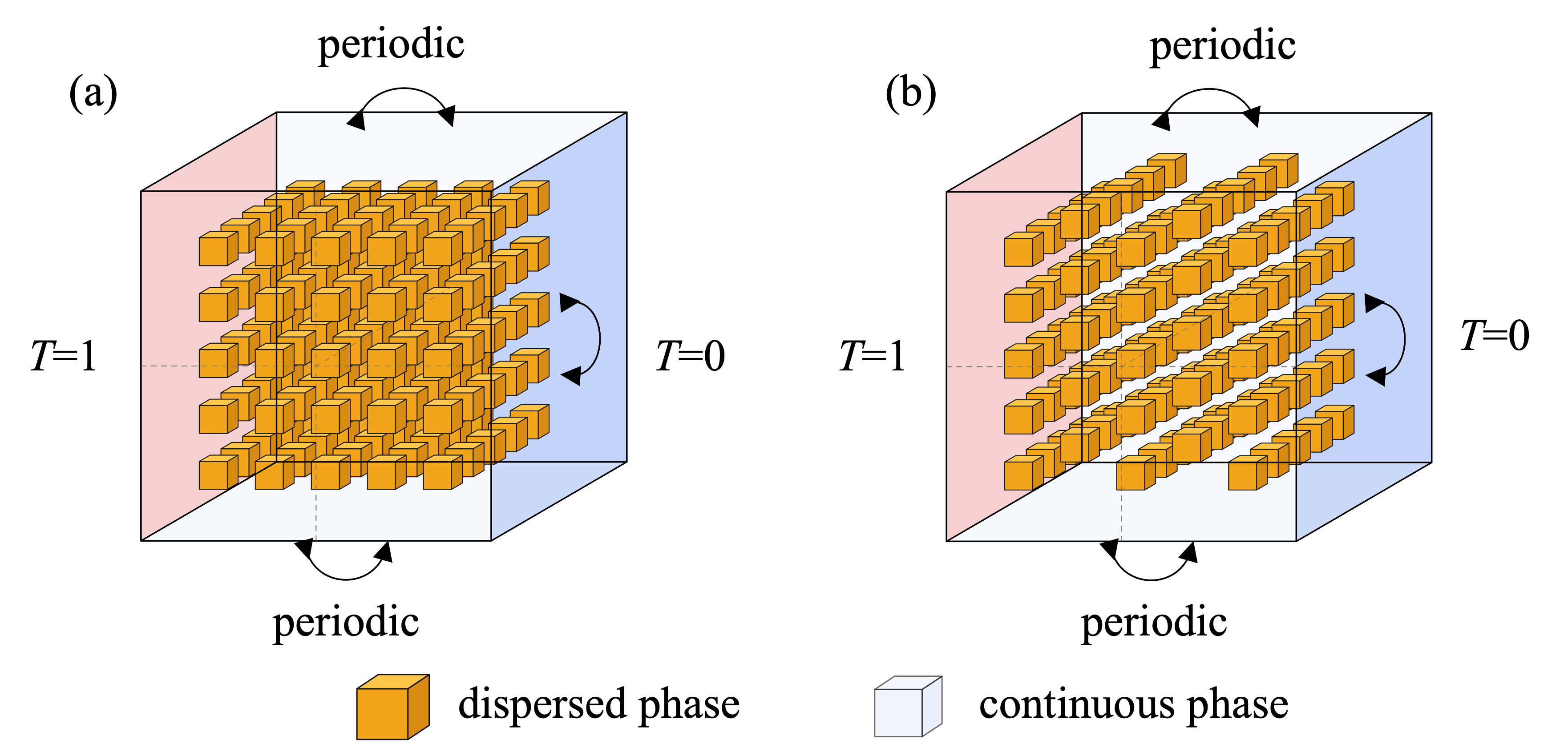}
    \caption{Anisotropic porous cube-array geometries used for the effective-conductivity measurement. Panel (a) shows the ordered array, where the dispersed cubes remain aligned through successive streamwise layers. Panel (b) shows the cross array, where alternating layers are shifted in the transverse plane. The dispersed cube phase is embedded in a continuous conducting matrix between the hot and cold plates, and the transverse directions are periodic.}
    \label{fig:cube_geometries}
\end{figure}

Anisotropic porous materials and architected composites arise in thermal barrier systems, porous heat exchangers, battery electrodes, and geomaterials, where the apparent heat transport depends on both tensor orientation and microstructural connectivity. Effective-conductivity prediction in such media must account for stochastic phase geometry, tortuous conducting pathways, and conjugate heat transfer across numerous phase interfaces~\citep{WangWangPanChen2007PRE,WangPan2008IJHMT,WangPan2008MSER,WangKangPan2009ATE}. We therefore measure steady heat conduction through two periodic cube-array porous media (Fig.~\ref{fig:cube_geometries}): the ordered array in Fig.~\ref{fig:cube_geometries}(a) aligns the dispersed cubes from layer to layer, while the cross array in Fig.~\ref{fig:cube_geometries}(b) staggers successive layers in the transverse plane at the same volume fraction.

Both calculations use a $128^3$ domain with cubes of side length 12 lattice units and a dispersed-phase volume fraction of $0.103$. The continuous matrix and dispersed cubes are assigned the principal conductivities
\begin{equation}
\boldsymbol{k}_c=(100,100,0.01),\qquad
\boldsymbol{k}_d=(10,10,0.01)
\quad \mathrm{W\,m^{-1}K^{-1}},
\label{eq:app-cube-principal-k}
\end{equation}
respectively. In each run, the two phases share the same ZYZ Euler angles, so the comparison isolates the combined effect of phase arrangement and tensor orientation. The angle set is
\begin{equation}
(\alpha_E,\beta_E,\gamma_E)
\in
\{(105^\circ,50^\circ,25^\circ),
(115^\circ,50^\circ,25^\circ),
(125^\circ,50^\circ,25^\circ),
(135^\circ,50^\circ,25^\circ)\}.
\label{eq:app-cube-euler-set}
\end{equation}
Since the first two principal conductivities are equal in both phases, the local tensor is axisymmetric about the low-conductivity principal direction. The sweep in Eq.~\eqref{eq:app-cube-euler-set} therefore mainly changes the projection of this low-conductivity axis onto the imposed temperature-gradient direction.

The effective coefficients are obtained from the conserved LB heat flux, which ties the reported macroscopic response to the physical flux carried by the populations. At steady state, the non-equilibrium first moment is
\begin{equation}
\mathbf{j}^{\mathrm{neq}}
=
\sum_i \mathbf{c}_i
\left(g_i-g_i^{\mathrm{eq}}\right),
\end{equation}
and the heat flux used in the measurement is evaluated as
\begin{equation}
\mathbf{q}
=
\mathbf{j}^{\mathrm{neq}}
-\frac{1}{2}\mathbf{S}\mathbf{j}^{\mathrm{neq}},
\label{eq:app-cube-lb-flux}
\end{equation}
where $\mathbf{S}$ is the local first-order relaxation tensor associated with the rotated conductivity. The reported heat-flux density is the plateau value of the cross-section-averaged $\mathbf{q}$ in the interior of the sample. With $G_x=\Delta T/L_x$, the measured effective tensor entries are
\begin{equation}
K_{\ell x}^{\mathrm{eff}}
=
\frac{\bar q_\ell}{G_x},
\qquad \ell\in\{x,y,z\}.
\label{eq:app-cube-effective-k}
\end{equation}
For consistency with the symmetric conductivity notation used above, the transverse entries measured from $\bar q_y$ and $\bar q_z$ are reported as $K_{xy}^{\mathrm{eff}}$ and $K_{xz}^{\mathrm{eff}}$, respectively.

\begin{table}[pos=H]
    \centering
    \rmfamily
    \caption{Plateau heat fluxes and effective conductivity entries for the anisotropic porous cube arrays. Heat fluxes are in $\mathrm{W\,m^{-2}}$ for $G_x=1/128~\mathrm{K\,m^{-1}}$, and conductivities are in $\mathrm{W\,m^{-1}K^{-1}}$.}
    \label{tab:cube-effective-conductivity}
    \begingroup
    \small
    \setlength{\tabcolsep}{2.6pt}
    \renewcommand{\arraystretch}{1.08}
    \begin{tabular*}{\textwidth}{@{\extracolsep{\fill}}llrrrrrr@{}}
        \toprule
        Geometry &
        $(\alpha_E,\beta_E,\gamma_E)$ &
        $\bar q_x$ &
        $\bar q_y$ &
        $\bar q_z$ &
        $K_{xx}^{\mathrm{eff}}$ &
        $K_{xy}^{\mathrm{eff}}$ &
        $K_{xz}^{\mathrm{eff}}$ \\
        \midrule
        Ordered & $(105^\circ,50^\circ,25^\circ)$ & 0.4733 & 0.0723 & 0.0628 & 60.58 & 9.25 & 8.04 \\
        Ordered & $(115^\circ,50^\circ,25^\circ)$ & 0.4382 & 0.1098 & 0.1022 & 56.09 & 14.05 & 13.08 \\
        Ordered & $(125^\circ,50^\circ,25^\circ)$ & 0.3928 & 0.1344 & 0.1374 & 50.28 & 17.20 & 17.59 \\
        Ordered & $(135^\circ,50^\circ,25^\circ)$ & 0.3426 & 0.1433 & 0.1680 & 43.85 & 18.34 & 21.51 \\
        \midrule
        Cross & $(105^\circ,50^\circ,25^\circ)$ & 0.4636 & 0.0714 & 0.0608 & 59.34 & 9.14 & 7.78 \\
        Cross & $(115^\circ,50^\circ,25^\circ)$ & 0.4321 & 0.1087 & 0.1003 & 55.30 & 13.91 & 12.84 \\
        Cross & $(125^\circ,50^\circ,25^\circ)$ & 0.3898 & 0.1332 & 0.1364 & 49.89 & 17.05 & 17.47 \\
        Cross & $(135^\circ,50^\circ,25^\circ)$ & 0.3416 & 0.1425 & 0.1679 & 43.73 & 18.23 & 21.50 \\
        \bottomrule
    \end{tabular*}
    \endgroup
\end{table}

Table~\ref{tab:cube-effective-conductivity} shows a clear orientational response. As $\alpha_E$ increases from $105^\circ$ to $135^\circ$, the effective streamwise conductivity decreases from $60.58$ to $43.85~\mathrm{W\,m^{-1}K^{-1}}$ in the ordered array and from $59.34$ to $43.73~\mathrm{W\,m^{-1}K^{-1}}$ in the cross array. The corresponding reductions are about $27.6\%$ and $26.3\%$. This trend follows directly from the tensor rotation: at the larger values of $\alpha_E$, the low-conductivity principal direction has a stronger projection onto the imposed $x$-direction gradient, which reduces the streamwise component of the macroscopic flux.

The same rotation also changes the direction of the heat flux. The transverse coefficients increase monotonically over the sweep. In the ordered array, $K_{xy}^{\mathrm{eff}}$ rises from $9.25$ to $18.34~\mathrm{W\,m^{-1}K^{-1}}$, while $K_{xz}^{\mathrm{eff}}$ rises from $8.04$ to $21.51~\mathrm{W\,m^{-1}K^{-1}}$. The cross array gives nearly the same transverse response, with final values $18.23$ and $21.50~\mathrm{W\,m^{-1}K^{-1}}$. Thus, a single imposed temperature gradient generates a progressively more oblique heat-flux vector as the material axes are rotated. Under periodic transverse boundaries, these values are periodic-volume averages of the tensor-induced flux components associated with the imposed $x$-direction gradient.

\begin{figure}[pos=H]
    \centering
    \includegraphics[width=0.98\textwidth]{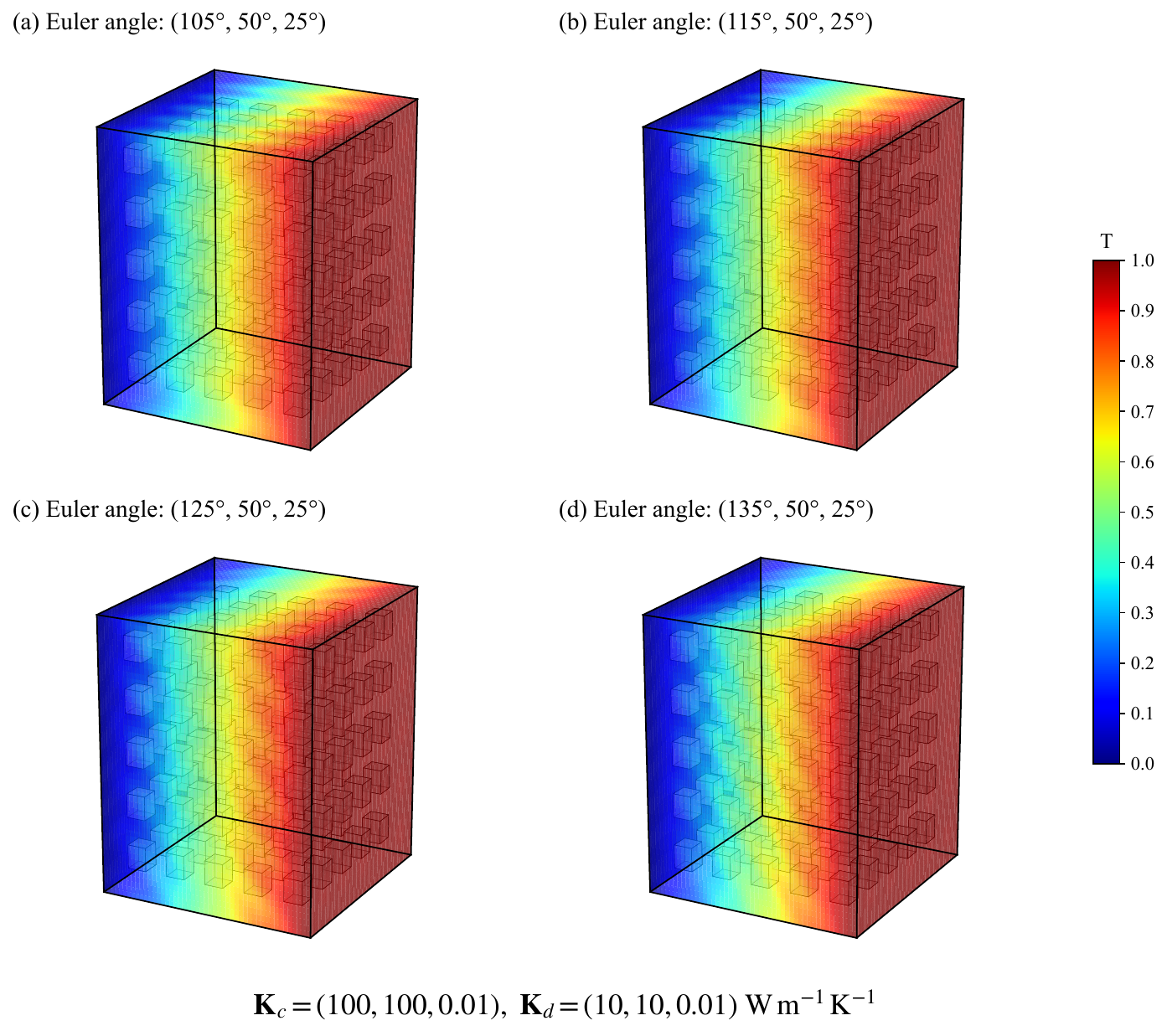}
    \caption{Temperature distributions in the ordered cube array for the four Euler-angle cases in Table~\ref{tab:cube-effective-conductivity}: (a) $(105^\circ,50^\circ,25^\circ)$, (b) $(115^\circ,50^\circ,25^\circ)$, (c) $(125^\circ,50^\circ,25^\circ)$, and (d) $(135^\circ,50^\circ,25^\circ)$. The continuous and dispersed phases use $\boldsymbol{k}_c=(100,100,0.01)$ and $\boldsymbol{k}_d=(10,10,0.01)~\mathrm{W\,m^{-1}K^{-1}}$, respectively.}
    \label{fig:ordered_cubes_temperature}
\end{figure}

\begin{figure}[pos=H]
    \centering
    \includegraphics[width=0.98\textwidth]{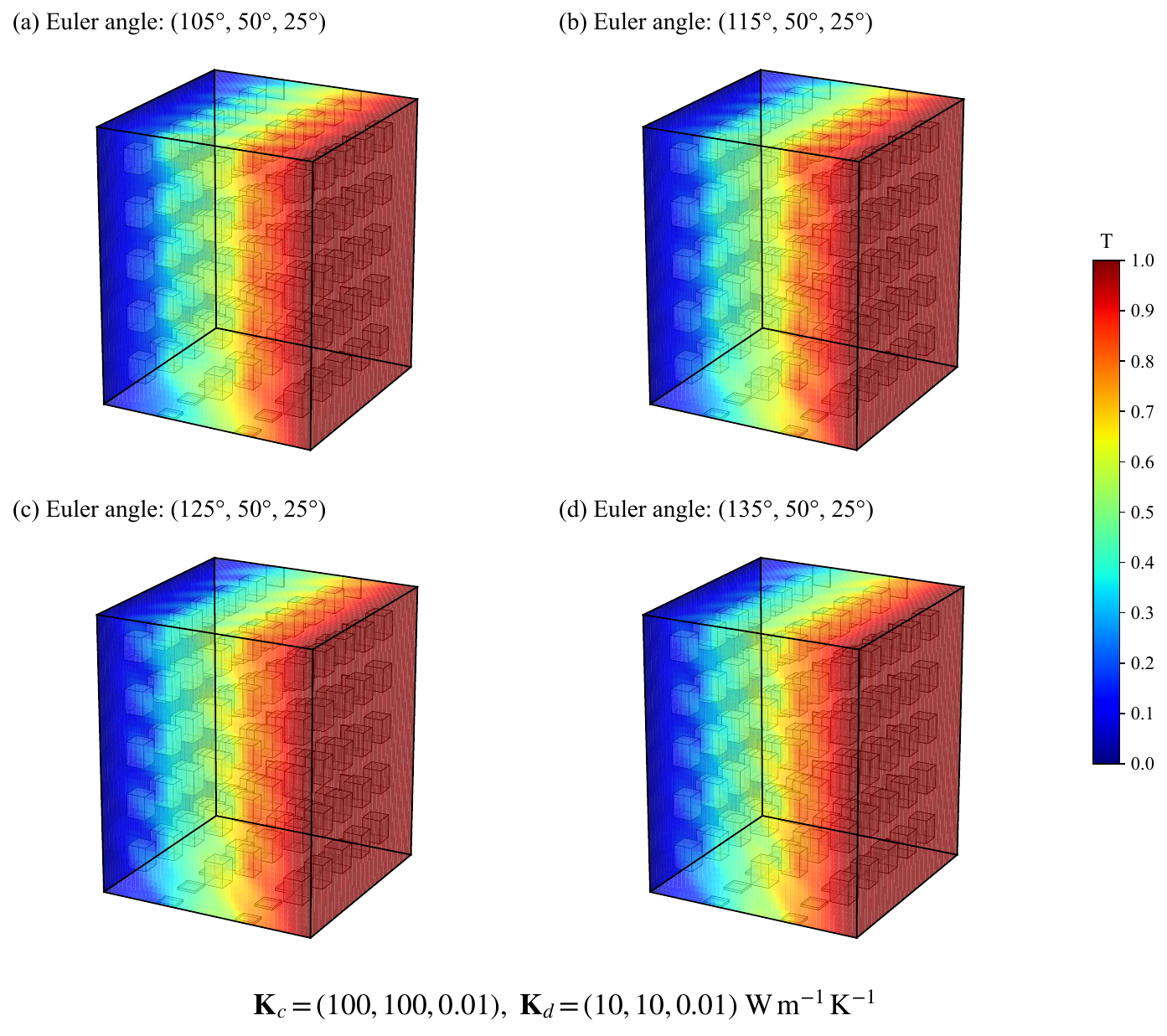}
    \caption{Temperature distributions in the cross cube array for the same material tensors as Fig.~\ref{fig:ordered_cubes_temperature}: (a) $(105^\circ,50^\circ,25^\circ)$, (b) $(115^\circ,50^\circ,25^\circ)$, (c) $(125^\circ,50^\circ,25^\circ)$, and (d) $(135^\circ,50^\circ,25^\circ)$. The staggered arrangement changes the local thermal path through the cube layers while maintaining the same phase volume fraction.}
    \label{fig:cross_cubes_temperature}
\end{figure}

The temperature fields in Figs.~\ref{fig:ordered_cubes_temperature} and~\ref{fig:cross_cubes_temperature} provide the spatial interpretation of the effective coefficients. In all cases, the dominant temperature variation remains aligned with the imposed hot--cold direction, but the isothermal surfaces are locally deflected around the dispersed cubes. These deflections are strongest near the cube interfaces, where both the phase contrast and the off-diagonal tensor entries influence the local heat path. The cross arrangement produces a slightly more tortuous thermal path at the smaller Euler angles, which is reflected in the lower $K_{xx}^{\mathrm{eff}}$ values relative to the ordered array. The difference between the two structures decreases from about $2.1\%$ at $(105^\circ,50^\circ,25^\circ)$ to below $0.3\%$ at $(135^\circ,50^\circ,25^\circ)$, indicating that the orientation of the anisotropic tensor becomes the dominant control on the streamwise conductance at the larger angles.

Together, the two cube-array calculations demonstrate that the present ELBM can measure the effective thermal response of three-dimensional porous materials with strongly rotated anisotropic phases. The method captures both the reduction of the imposed-direction conductivity and the emergence of transverse effective tensor entries from the same steady calculation, using the physical heat flux carried by the LB populations.

\subsection{Anisotropic Rayleigh--B\'enard convection}
\label{sec:application-anisotropic-rbc}

Anisotropic thermal convection arises in fluid layers where the momentum balance can be approximated by an incompressible Newtonian flow, while heat diffuses preferentially along a material direction. A direct laboratory example is a nematic liquid crystal with an externally aligned or strongly anchored director: in 4-cyano-4'-pentylbiphenyl (5CB), the thermal conductivity measured parallel to the director differs from that measured perpendicular to it~\citep{AhlersEtAl1994ThermalConductivity5CB}, and Rayleigh--B\'enard studies show that director alignment can modify the convective heat flux~\citep{FengPeschKramer1992NematicRBC,WeissAhlers2013NematicRBC}. Related tensorial heat-transport descriptions also appear after volume averaging in fluid-saturated porous and fibrous media, where thermal pathways are set by microstructural orientation~\citep{NieldBejan2017PorousConvection}. The simplified case below therefore keeps the hydrodynamics at the incompressible Newtonian level and isolates how a prescribed thermal-diffusion tensor changes plume morphology and heat transport.

The preceding examples prescribe the advecting velocity or measure a steady conductive response. We now consider a buoyancy-driven problem in which the velocity and temperature fields are coupled dynamically. This case probes the anisotropic scalar solver in a feedback loop: anisotropic thermal diffusion changes plume formation, the plume field modifies the buoyancy force, and the resulting flow sets the vertical heat flux. The problem also provides a direct physical diagnostic of the tensorial diffusion model, since changing the horizontal diffusivity alters the lateral plume scale before it changes the imposed vertical conductive reference flux.

The calculation is a two-dimensional Rayleigh--B\'enard system in a domain of height $H$ and horizontal period $L_x=2H$. The bottom plate is held at the hot temperature $\theta_h=1$, the top plate at the cold temperature $\theta_c=0$, and the horizontal direction is periodic. The velocity satisfies no-slip boundary conditions at the plates. In the Boussinesq approximation, the governing equations are
\begin{subequations}
\label{eq:app-rbc-governing}
\begin{align}
\nabla\cdot\mathbf{u} &= 0,
\label{eq:app-rbc-continuity}
\\
\partial_t\mathbf{u}
 + \mathbf{u}\cdot\nabla\mathbf{u}
&=
-\nabla p
 + \nu\nabla^2\mathbf{u}
 + g\beta\left(\theta-\theta_{\mathrm{cond}}\right)\mathbf{e}_z,
\label{eq:app-rbc-ns}
\\
\partial_t\theta
 + \nabla\cdot(\mathbf{u}\theta)
&=
\nabla\cdot\left(\mathbf{D}\nabla\theta\right),
\label{eq:app-rbc-ade}
\end{align}
\end{subequations}
where $\mathbf{u}=(u_x,u_z)$ is the velocity, $p$ is the kinematic pressure, $\nu$ is the kinematic viscosity, $g$ is the gravitational acceleration, $\beta$ is the thermal expansion coefficient, $\mathbf{e}_z$ is the upward unit vector, and $\theta$ is the temperature. The conductive reference profile is
\begin{equation}
\theta_{\mathrm{cond}}(z)
=
\theta_h-\Delta\theta\,\frac{z}{H},
\qquad
\Delta\theta=\theta_h-\theta_c .
\label{eq:app-rbc-conduction-profile}
\end{equation}
The buoyancy force is written in terms of the departure from the conductive reference state, so the linear conductive solution remains a quiescent equilibrium before the initial perturbation grows. The thermal diffusivity tensor is aligned with the coordinate axes,
\begin{equation}
\mathbf{D}
=
\begin{bmatrix}
d_1 & 0\\
0 & d_2
\end{bmatrix},
\qquad
r=\frac{d_2}{d_1},
\label{eq:app-rbc-diffusion-tensor}
\end{equation}
with $d_1$ the horizontal diffusivity, $d_2$ the vertical diffusivity, and $\kappa_v$ the reference vertical thermal diffusivity. The reported sweep keeps the vertical diffusivity fixed at $d_2=\kappa_v=3.0\times10^{-3}$ and varies $d_1=d_2/r$. This choice isolates lateral thermal smoothing: the conductive reference state, the imposed vertical diffusive flux, and the Rayleigh number defined below are common to all cases, and the horizontal dissipation of temperature anomalies is the swept quantity. Small $r$ corresponds to strong horizontal thermal diffusion. Large $r$ corresponds to weak horizontal diffusion and a stronger tendency to form narrow horizontal thermal structures.

The Rayleigh number and vertical Prandtl number are defined using the vertical diffusivity,
\begin{equation}
Ra
=
\frac{g\beta\Delta\theta H^3}{\nu\kappa_v},
\qquad
Pr_v
=
\frac{\nu}{\kappa_v}.
\label{eq:app-rbc-parameters}
\end{equation}
All cases use $Ra=10^9$, $Pr_v=1$, $\nu=3.0\times10^{-3}$, and $H=1024$ lattice units, giving a $2048\times1024$ domain. The value $Ra=10^9$ places the calculation in a high-Rayleigh-number regime with thin thermal boundary layers, intermittent plume emission, and a strongly convective heat flux. The initial temperature field is the conductive profile plus a random perturbation of amplitude $10^{-3}$. The stable sweep covers
\begin{equation}
r\in\{10^{-4},10^{-3},10^{-2},10^{-1},1,10,10^2,10^3\}.
\label{eq:app-rbc-rset}
\end{equation}

The vertical heat flux used in the analysis is
\begin{equation}
q_z(x,z)
=
u_z\theta
-\left(\mathbf{D}\nabla\theta\right)_z ,
\label{eq:app-rbc-heat-flux}
\end{equation}
and the conductive reference flux is $q_0=\kappa_v\Delta\theta/H$. We report the bulk Nusselt number
\begin{equation}
Nu_{\mathrm{bulk}}
=
1+
\frac{H}{\kappa_v\Delta\theta}
\left\langle u_z\theta\right\rangle_{x,z},
\label{eq:app-rbc-nu-bulk}
\end{equation}
where $\langle\cdot\rangle_{x,z}$ denotes a volume average over the full domain. A second estimate, $Nu_{\mathrm{flux}}$, is obtained from the interior plateau of $\langle q_z\rangle_x/q_0$, with $\langle\cdot\rangle_x$ denoting a horizontal average at fixed height. The agreement between these two independent estimates is used below as the primary heat-flux closure check.

\begin{figure}[pos=H]
    \centering
    \includegraphics[width=0.98\textwidth]{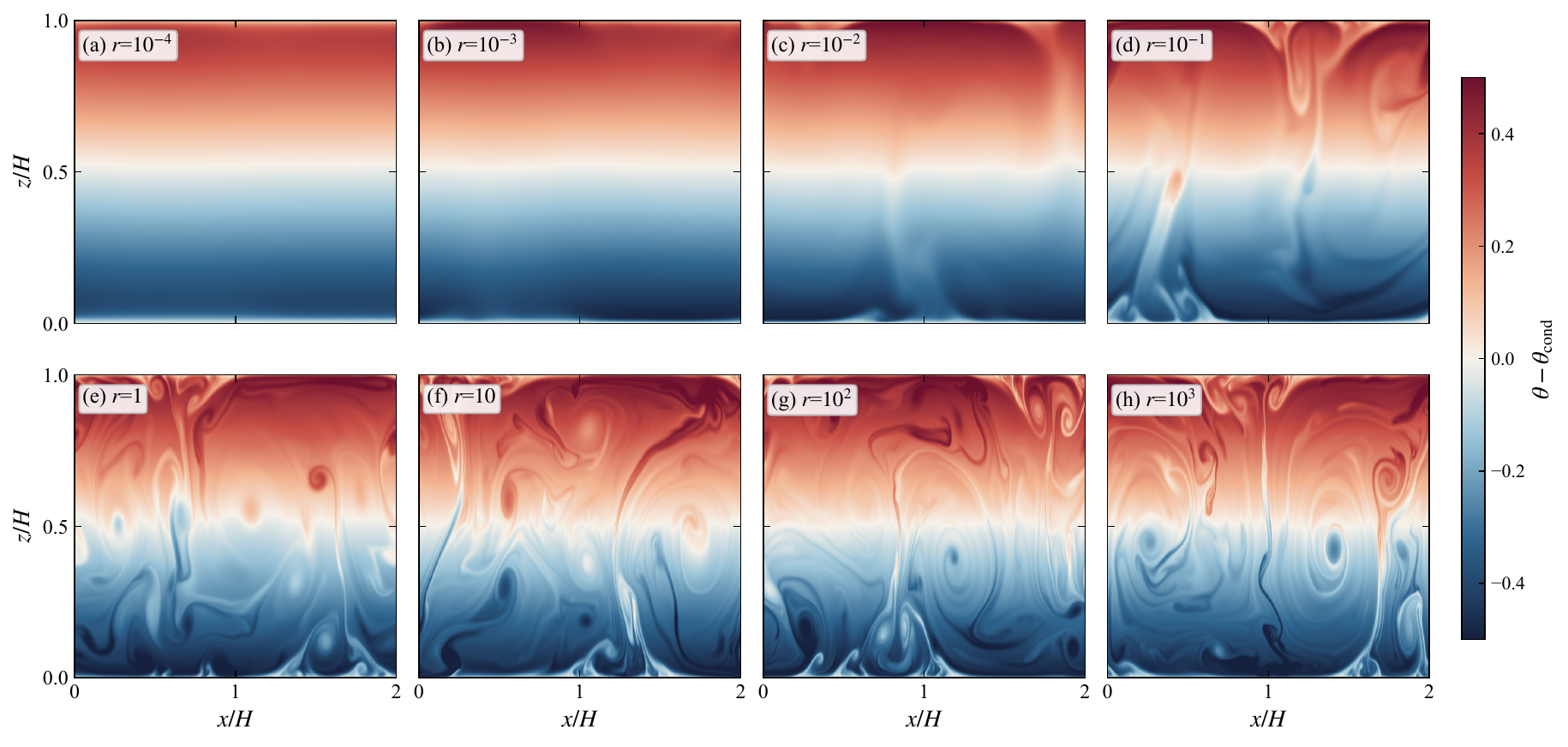}
    \caption{Temperature departures from the conductive profile in anisotropic Rayleigh--B\'enard convection at $Ra=10^9$ and $Pr_v=1$. Panels (a)--(h) show increasing values of $r=d_2/d_1$ from $10^{-4}$ to $10^3$. The vertical diffusivity $d_2$ is fixed in all cases, while the horizontal diffusivity decreases as $r$ increases.}
    \label{fig:rbc_temperature_anomaly_grid}
\end{figure}

Figure~\ref{fig:rbc_temperature_anomaly_grid} shows the late-time temperature anomaly $\theta-\theta_{\mathrm{cond}}$ for the eight stable anisotropy ratios. At $r=10^{-4}$ and $10^{-3}$ the field is strongly smoothed in the horizontal direction. The anomaly is dominated by broad layers and only weakly deformed plume fronts near the plates, because the large $d_1$ damps lateral temperature gradients before narrow plume roots can persist. Once $r$ reaches $10^{-2}$ and $10^{-1}$, coherent upwelling and downwelling structures become visible, with plume roots forming at both thermal boundary layers. The isotropic case $r=1$ already contains a richer population of vertical thermal sheets and rolled-up interfaces. Increasing $r$ further produces finer, more filamentary structures, especially near the hot and cold plates where plumes are generated. The sequence therefore gives a direct spatial picture of the role of horizontal diffusion: reducing $d_1$ sharpens lateral temperature gradients, shifts the thermal field toward higher horizontal wavenumbers, and increases the spatial intermittency of plume-mediated heat transport.

\begin{figure}[pos=H]
    \centering
    \includegraphics[width=0.98\textwidth]{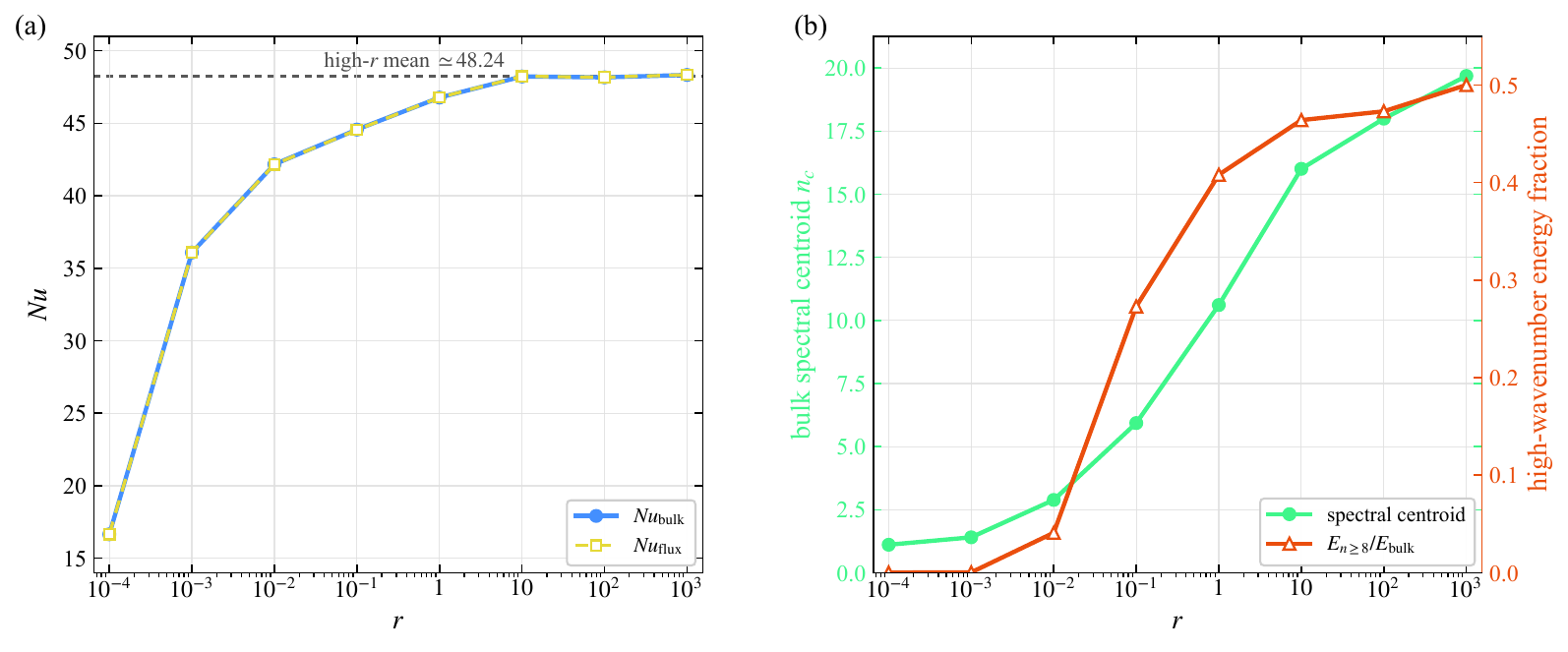}
    \caption{Heat transport and plume morphology as functions of $r=d_2/d_1$. Panel (a) compares the bulk Nusselt number with the flux-plateau estimate. Panel (b) reports two bulk plume statistics computed from $\theta-\theta_{\mathrm{cond}}$ over $0.2H\le z\le0.8H$: the horizontal spectral centroid $n_c$ and the fraction of spectral energy in modes $n\ge8$.}
    \label{fig:rbc_nusselt_plume_metrics}
\end{figure}

Figure~\ref{fig:rbc_nusselt_plume_metrics} separates the heat-transfer response from the normalized bulk morphology metrics. The heat transport rises sharply when $r$ is increased from $10^{-4}$ to $10^{-3}$: $Nu_{\mathrm{bulk}}$ grows from $16.65$ to $36.08$. This first jump should not be interpreted as an immediate transfer of temperature-anomaly energy to high horizontal wavenumbers. Instead, the two lowest-$r$ cases remain dominated by large-scale horizontal modes, while the weaker horizontal diffusion at $r=10^{-3}$ permits a much stronger plume-mediated heat flux and a larger positive correlation between $u_z$ and $\theta$. Further increases in $r$ continue to enhance the heat flux with a decreasing slope. By $r=10$, the Nusselt number has reached $48.22$, and the cases $r=10$, $10^2$, and $10^3$ remain clustered around $Nu\simeq48.2$--$48.3$.

The morphology statistics quantify the subsequent shift in plume scale rather than the total anomaly amplitude. Writing
\(\theta'=\theta-\theta_{\mathrm{cond}}\), the horizontal mean \(\langle\theta'\rangle_x(z)\) is first removed at each height in the bulk band \(0.2H\le z\le0.8H\). The horizontal Fourier coefficients and modal energy are then defined as
\begin{equation}
\widehat{\theta}_n(z)
=
\frac{1}{L_x}\int_0^{L_x}
\left[\theta'(x,z)-\langle\theta'\rangle_x(z)\right]
\exp\!\left(-\mathrm{i}\frac{2\pi n x}{L_x}\right)\,dx,
\qquad
E_n=\left\langle |\widehat{\theta}_n(z)|^2\right\rangle_{0.2H\le z\le0.8H}.
\label{eq:app-rbc-fourier-energy}
\end{equation}
Here $n$ is the horizontal Fourier mode number, $\mathrm{i}^2=-1$, and the brackets in $E_n$ denote an average over the stated bulk band. The spectral centroid
\begin{equation}
n_c
=
\frac{\sum_{n\ge1}nE_n}{\sum_{n\ge1}E_n},
\label{eq:app-rbc-spectral-centroid}
\end{equation}
increases from $1.11$ at $r=10^{-4}$ to $19.68$ at $r=10^3$. Over the first decade, however, it changes only from $1.11$ to $1.41$, and the high-wavenumber fraction $\sum_{n\ge8}E_n/\sum_{n\ge1}E_n$ remains near zero ($2.0\times10^{-4}$ to $4.2\times10^{-4}$). The substantial transfer of normalized bulk energy to smaller horizontal scales begins only from $r\simeq10^{-2}$ onward, and the high-wavenumber fraction reaches about $0.50$ by $r=10^3$. These statistics therefore confirm the fine-scale content of the high-$r$ cases, while the initial Nusselt-number rise mainly reflects stronger heat-flux organization in still large-scale structures.

The heat-transfer curve reaches a plateau over the same high-$r$ range. Reducing $d_1$ continues to sharpen lateral temperature gradients; the vertical diffusivity $d_2$ and the buoyant driving contained in $Ra$ remain fixed. Once horizontal smoothing no longer limits plume survival, further narrowing of the plume filaments supplies little additional vertical heat transport. The observed cluster at $Nu\simeq48.2$--$48.3$ therefore marks the high-$r$ heat-transfer capacity of this fixed-\(Ra\), fixed-\(Pr_v\) configuration, controlled by thermal-boundary-layer exchange and the induced circulation.

\begin{figure}[pos=H]
    \centering
    \includegraphics[width=0.98\textwidth]{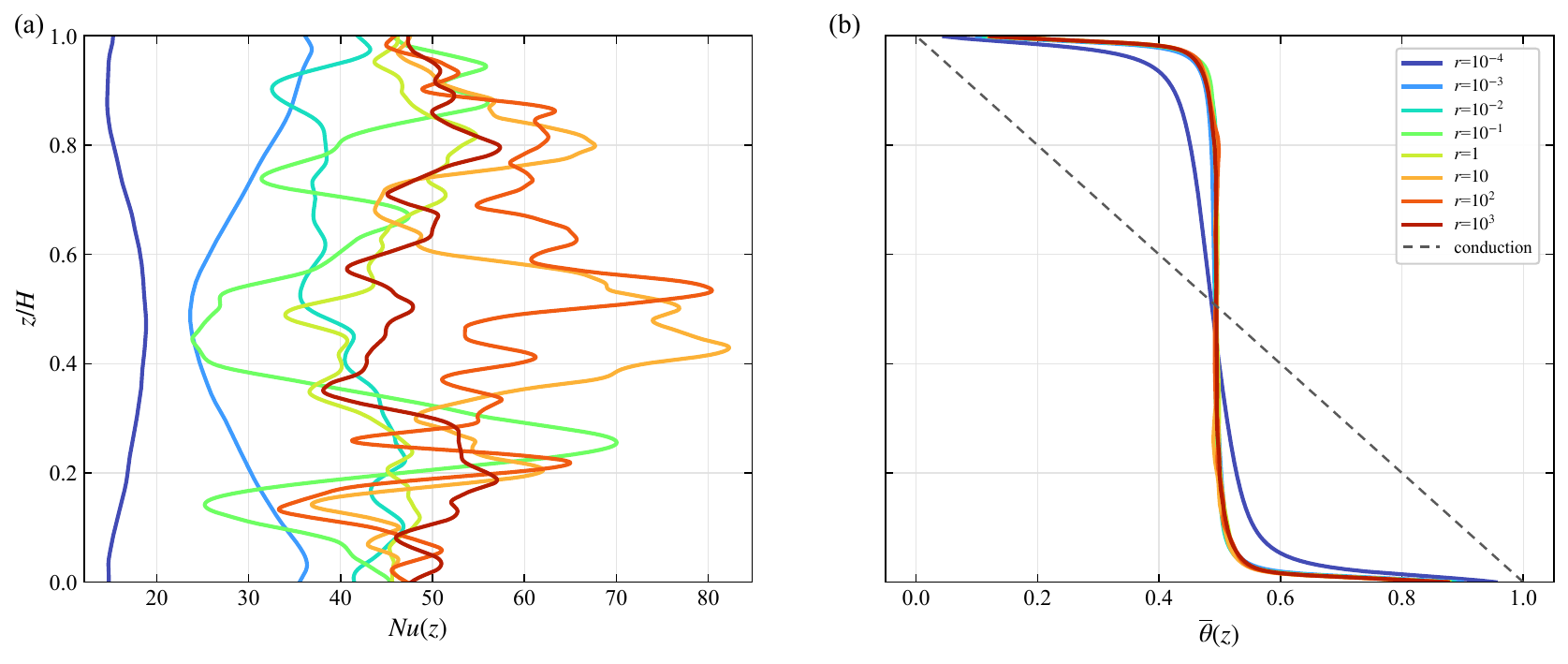}
    \caption{Late-time vertical profiles for the anisotropic Rayleigh--B\'enard sweep. Panel (a) shows the cross-section-averaged heat-flux profile $Nu(z)=\langle q_z\rangle_x/q_0$. Panel (b) shows the horizontally averaged temperature profile, with the dashed line denoting the conductive solution.}
    \label{fig:rbc_flux_temperature_profiles}
\end{figure}

Figure~\ref{fig:rbc_flux_temperature_profiles} connects the global Nusselt numbers to the vertical structure of the layer. The heat-flux profiles occupy distinct levels at small and moderate $r$, consistent with the increase in $Nu$ shown in Fig.~\ref{fig:rbc_nusselt_plume_metrics}(a). The $r=10^{-4}$ profile remains far below the other cases through most of the layer, reflecting the strong horizontal smoothing seen in Fig.~\ref{fig:rbc_temperature_anomaly_grid}(a). The profiles for $r\ge1$ lie much closer together, with fluctuations caused by the finite-time plume organization and comparable bulk levels. The temperature profiles in Fig.~\ref{fig:rbc_flux_temperature_profiles}(b) show the corresponding convective state: most of the layer is well mixed around $\bar{\theta}\approx0.5$, with steep gradients confined to thin regions near the plates. The low-$r$ case retains visibly thicker thermal transition regions, consistent with its smaller Nusselt number.
The flow diagnostics reported in Table~\ref{tab:rbc-nusselt-closure} are
\begin{equation}
u_{\mathrm{rms}}
=
\left\langle u_x^2+u_z^2\right\rangle_{x,z}^{1/2},
\qquad
\|\nabla\cdot\mathbf{u}\|_2
=
\left\langle \left(\nabla\cdot\mathbf{u}\right)^2\right\rangle_{x,z}^{1/2}.
\label{eq:app-rbc-flow-diagnostics}
\end{equation}

\begin{table}[pos=H]
    \centering
    \rmfamily
    \caption{Nusselt-number closure and flow diagnostics for the stable anisotropic Rayleigh--B\'enard sweep. $Nu_{\mathrm{bulk}}$ is the volume estimate in Eq.~\eqref{eq:app-rbc-nu-bulk}, while $Nu_{\mathrm{flux}}$ is obtained from the interior plateau of $\langle q_z\rangle_x/q_0$. The relative difference is $\lvert Nu_{\mathrm{flux}}-Nu_{\mathrm{bulk}}\rvert/Nu_{\mathrm{bulk}}$.}
    \label{tab:rbc-nusselt-closure}
    \begingroup
    \small
    \setlength{\tabcolsep}{4.2pt}
    \renewcommand{\arraystretch}{1.08}
    \begin{tabular*}{\textwidth}{@{\extracolsep{\fill}}lrrrrr@{}}
        \toprule
        $r=d_2/d_1$ & $Nu_{\mathrm{bulk}}$ & $Nu_{\mathrm{flux}}$ & rel. diff. (\%) & $u_{\mathrm{rms}}$ & $\|\nabla\cdot\mathbf{u}\|_2$ \\
        \midrule
        $10^{-4}$ & 16.654 & 16.653 & 0.009 & 0.0292 & $3.61\times10^{-6}$ \\
        $10^{-3}$ & 36.085 & 36.086 & 0.004 & 0.0392 & $6.91\times10^{-6}$ \\
        $10^{-2}$ & 42.165 & 42.158 & 0.018 & 0.0413 & $7.92\times10^{-6}$ \\
        $10^{-1}$ & 44.564 & 44.544 & 0.044 & 0.0419 & $8.14\times10^{-6}$ \\
        $1$ & 46.768 & 46.775 & 0.014 & 0.0422 & $8.39\times10^{-6}$ \\
        $10$ & 48.223 & 48.235 & 0.026 & 0.0403 & $8.57\times10^{-6}$ \\
        $10^2$ & 48.163 & 48.172 & 0.018 & 0.0423 & $8.84\times10^{-6}$ \\
        $10^3$ & 48.329 & 48.343 & 0.029 & 0.0412 & $8.51\times10^{-6}$ \\
        \bottomrule
    \end{tabular*}
    \endgroup
\end{table}

Table~\ref{tab:rbc-nusselt-closure} summarizes the heat-transfer measurements using the two Nusselt estimates that are defined identically across the full archived sweep. The bulk and flux-plateau estimates agree within $4.4\times10^{-2}\%$ for all stable cases. The isotropic baseline, $r=1$, gives $Nu_{\mathrm{bulk}}=46.77$ at $Ra=10^9$ and $Pr_v=1$, within about $3\%$ of an independent two-dimensional direct numerical simulation result at the same Rayleigh and Prandtl numbers \citep{ZhuEtAl2018UltimateRBC}. The root-mean-square velocity remains at $O(10^{-2})$, and the divergence diagnostic stays below $9\times10^{-6}$, so the reported Nusselt trend is supported by a consistent heat-flux balance in a nearly incompressible flow field.

The combined physical picture is governed by lateral smoothing at fixed vertical diffusivity and fixed $Ra$. Strong horizontal diffusion erases temperature anomalies before they develop into narrow rising and falling structures, weakens the convective correlation $\langle u_z\theta\rangle$, and produces low Nusselt numbers. As the horizontal diffusivity is reduced, the plume field gains high-wavenumber content and the heat transfer increases rapidly. At large $r$, the plume morphology continues to sharpen while the Nusselt number approaches a plateau. This high-anisotropy regime is controlled by vertical exchange through the thermal boundary layers and the induced circulation; further reduction of horizontal diffusivity mainly changes the fine structure of the temperature field.

%% file: sections/conclusion.tex
\section{Conclusions}
\label{sec:conclusion}

This paper developed a local entropic lattice Boltzmann formulation for the full-tensor anisotropic advection--diffusion equation. The method separates the non-equilibrium population into a first-order flux sector and a residual ghost sector. The prescribed diffusion tensor is imposed directly through a local tensorial relaxation of the non-equilibrium flux, while an ADE-corrected entropic relaxation damps the remaining kinetic content. A geometric positivity fallback preserves the transported scalar when the full collision step would leave the positive branch. Chapman--Enskog analysis shows that this construction recovers the target ADE with the discrete-time tensor diffusivity relation, and that cross-diffusion is produced by the same flux-space matrix action that represents the physical tensor.

The validation cases confirm this tensor recovery across constant, rotated, spatially varying, advective, and source-driven transport. The Gaussian-plume and sinusoidal-decay tests recover principal-axis spreading, off-diagonal diffusion, and projected decay rates for anisotropy ratios up to $10^4:10:1$. The variable-coefficient benchmark extends the verification to high-P\'eclet anisotropic advection--diffusion with local tensor contrasts up to $3\times10^4:1$, while retaining errors at the $10^{-3}$ level or below. These results support the intended role of the scheme as a local, matrix-free solver for strongly anisotropic ADEs.

The application studies show that the same collision structure applies when the tensor has a physical origin. For elongated Brownian rods, the method reproduces the long-time transport generated by orientation-dependent diffusion, cross-diffusion, and shear-induced migration. For heat conduction, it recovers rotated conductivity entries and off-diagonal heat fluxes in homogeneous solids, and captures how material-axis orientation and pore connectivity shape the effective response of anisotropic cube-array media. In anisotropic Rayleigh--B\'enard convection, reducing horizontal thermal smoothing sharpens plume structures and increases heat transport, demonstrating that the scalar solver can be embedded in a buoyancy-coupled flow calculation.

Several extensions follow naturally. In forced convection, the formulation can be used to study thermal or solutal transport in fibrous, laminated, architected, and additively manufactured materials. In natural convection, it provides a route to examine how directional conduction changes plume emission, boundary-layer exchange, and heat-transfer scaling in anisotropic porous media, liquid crystals, and aligned composites. Other targets include reactive or multicomponent transport in fractured media, battery electrodes, porous heat exchangers, and field-aligned transport in plasmas. Phase-change problems form another possible direction when latent heat, anisotropic conduction, and evolving permeability are represented within an enthalpy-based scalar transport model. Coupling the tensor field to orientation dynamics, microstructure evolution, or external alignment would further connect the method to materials whose transport axes evolve with the flow, temperature field, or processing conditions.

Overall, the formulation provides a local kinetic representation of anisotropic scalar transport in which the physical tensor is carried by the diffusive flux itself. It therefore offers a useful building block for simulations of direction-dependent transport in complex materials and coupled multiphysics systems.

%% file: sections/appendix.tex
\newcommand{\appendixsection}[1]{%
  \refstepcounter{section}%
  \section*{Appendix~\thesection. #1}%
  \addcontentsline{toc}{section}{Appendix~\thesection. #1}%
}

\appendixsection{Chapman--Enskog analysis for anisotropic advection--diffusion}
\label{app:ce-anisotropic-diffusion}

This appendix shows that the anisotropic ADE-KBC collision operator recovers the source-free form of Eq.~\eqref{eq:intro-ade} on the positive smooth branch where the geometric fallback is inactive. The derivation uses the tensor-product quadrature identities below; Greek indices denote Cartesian components, repeated Greek indices are summed, and $\delta_{\alpha\beta}$ is the Kronecker delta:
\begin{equation}
\sum_i W_i=1,\qquad
\sum_i W_i c_{i\alpha}=0,\qquad
\sum_i W_i c_{i\alpha}c_{i\beta}=c_s^2\delta_{\alpha\beta},\qquad
\sum_i W_i c_{i\alpha}c_{i\beta}c_{i\gamma}=0.
\end{equation}

\subsection{Multiscale expansion}

We adopt the diffusive scaling with formal ordering parameter $\epsilon$,
\begin{equation}
\nabla=\epsilon\nabla_1,\qquad
\partial_t=\epsilon\partial_{t_1}+\epsilon^2\partial_{t_2},\qquad
\mathbf{u}=\epsilon\mathbf{U},
\end{equation}
where $\mathbf{u}$ is the physical advecting velocity and $\mathbf{U}=O(1)$ is used only inside the expansion. The populations are written as
\begin{equation}
g_i=g_i^{(0)}+\epsilon g_i^{(1)}+\epsilon^2 g_i^{(2)}+O(\epsilon^3),
\end{equation}
with
\begin{equation}
\phi=\sum_i g_i^{(0)},\qquad \sum_i g_i^{(n)}=0\quad(n\ge 1).
\end{equation}
The entropic amplitude is expanded as
\begin{equation}
\lambda=\lambda^{(0)}+\epsilon\lambda^{(1)}+O(\epsilon^2).
\end{equation}

Under this scaling, the equilibrium distribution becomes
\begin{equation}
g_i^{eq}=g_i^{eq,(0)}+\epsilon g_i^{eq,(1)}+O(\epsilon^2),
\end{equation}
with
\begin{equation}
g_i^{eq,(0)}=W_i\phi,\qquad
g_i^{eq,(1)}=\frac{W_i}{c_s^2}\phi\,\mathbf{c}_i\cdot\mathbf{U}.
\end{equation}
The $O(\epsilon^2)$ equilibrium terms do not modify the leading ADE closure derived below. They are nevertheless retained in the implemented equilibrium because they reduce velocity-dependent truncation errors and enter the ADE-specific entropy correction.

The first-order non-equilibrium flux is
\begin{equation}
j_\alpha^{(1)}=\sum_i c_{i\alpha}\left(g_i^{(1)}-g_i^{eq,(1)}\right),
\end{equation}
and the first-order ghost part is
\begin{equation}
\Delta h_i^{(1)}
=\left(g_i^{(1)}-g_i^{eq,(1)}\right)-\mathcal{P}_i(\mathbf{j}^{(1)}).
\end{equation}
It satisfies
\begin{equation}
\sum_i\Delta h_i^{(1)}=0,\qquad
\sum_i c_{i\alpha}\Delta h_i^{(1)}=0.
\end{equation}

A second-order Taylor expansion of the streaming operator gives
\begin{equation}
\Delta t D_i g_i+\frac{\Delta t^2}{2}D_i^2 g_i=\Omega_i+O(\epsilon^3),\qquad
D_i=\partial_t+\mathbf{c}_i\cdot\nabla.
\end{equation}
Substitution of the multiscale expansion gives the order-by-order hierarchy used below; the superscripts on $\Omega_i^{(n)}$ denote the corresponding coefficients of the collision increment.

\subsection{Zeroth and first orders: tensorial Fick law}

At $O(\epsilon^0)$,
\begin{equation}
\Omega_i^{(0)}=0,
\end{equation}
so
\begin{equation}
g_i^{(0)}=g_i^{eq,(0)}=W_i\phi.
\end{equation}

At $O(\epsilon^1)$,
\begin{equation}
\Delta t D_{1i}g_i^{(0)}=\Omega_i^{(1)},\qquad
D_{1i}=\partial_{t_1}+\mathbf{c}_i\cdot\nabla_1,
\end{equation}
with
\begin{equation}
\Omega_i^{(1)}
=-\mathcal{P}_i(\mathbf{S}\mathbf{j}^{(1)})
-\lambda^{(0)}\Delta h_i^{(1)}.
\end{equation}
The zeroth moment gives
\begin{equation}
\partial_{t_1}\phi=0.
\end{equation}
The first moment gives
\begin{equation}
\Delta t\sum_i c_{i\alpha}c_{i\beta}\partial_{\beta_1}g_i^{(0)}
=-S_{\alpha\beta}j_\beta^{(1)},
\end{equation}
because the ghost part has zero first moment. Using $g_i^{(0)}=W_i\phi$ and the second-order quadrature identity,
\begin{equation}
c_s^2\Delta t\,\partial_{\alpha_1}\phi
=-S_{\alpha\beta}j_\beta^{(1)},
\end{equation}
and therefore
\begin{equation}
j_\alpha^{(1)}
=-c_s^2\Delta t\,(S^{-1})_{\alpha\beta}\partial_{\beta_1}\phi.
\label{eq:app-fick}
\end{equation}
This is the anisotropic Fick law generated by the first-order relaxation block. The entropic ghost amplitude does not enter because the ghost sector carries neither mass nor first-order flux.

\subsection{Second order: macroscopic closure}

At $O(\epsilon^2)$,
\begin{equation}
\Delta t\partial_{t_2}g_i^{(0)}
+\Delta t D_{1i}g_i^{(1)}
+\frac{\Delta t^2}{2}D_{1i}^2g_i^{(0)}
=\Omega_i^{(2)}.
\end{equation}
Taking the zeroth moment and using $\sum_i\Omega_i^{(2)}=0$ gives
\begin{equation}
\partial_{t_2}\phi
+\partial_{\alpha_1}\left(\sum_i c_{i\alpha}g_i^{(1)}\right)
+\frac{\Delta t}{2}\sum_i D_{1i}^2g_i^{(0)}=0.
\end{equation}
The first moment of $g_i^{(1)}$ contains the advective and diffusive contributions:
\begin{equation}
\sum_i c_{i\alpha}g_i^{(1)}
=\sum_i c_{i\alpha}g_i^{eq,(1)}+j_\alpha^{(1)}
=\phi U_\alpha+j_\alpha^{(1)}.
\end{equation}
Moreover, since $\partial_{t_1}\phi=0$ and the third-order velocity moment vanishes,
\begin{equation}
\sum_i D_{1i}^2g_i^{(0)}
=\sum_i c_{i\alpha}c_{i\beta}
\partial_{\alpha_1}\partial_{\beta_1}(W_i\phi)
=c_s^2\nabla_1^2\phi.
\end{equation}
Thus
\begin{equation}
\partial_{t_2}\phi+\nabla_1\cdot(\phi\mathbf{U})
+\nabla_1\cdot\mathbf{j}^{(1)}
+\frac{c_s^2\Delta t}{2}\nabla_1^2\phi=0.
\end{equation}
Substituting \eqref{eq:app-fick} yields
\begin{equation}
\partial_{t_2}\phi+\nabla_1\cdot(\phi\mathbf{U})
=\nabla_1\cdot\left[
c_s^2\Delta t\left(\mathbf{S}^{-1}-\frac{1}{2}\mathbf{I}\right)
\nabla_1\phi
\right].
\end{equation}
Returning to the physical variables with $\mathbf{u}=\epsilon\mathbf{U}$ gives the source-free form of Eq.~\eqref{eq:intro-ade}, with
\begin{equation}
\mathbf{D}=c_s^2\Delta t\left(\mathbf{S}^{-1}-\frac{1}{2}\mathbf{I}\right).
\label{eq:app-d-from-s}
\end{equation}
The tensorial relaxation matrix therefore determines the leading-order anisotropic diffusion tensor, while the scalar ghost relaxation affects only higher-order kinetic content on this branch. The principal-axis and cross-diffusion interpretation of this tensor is given in Sec.~\ref{sec:method-cross-diffusion}.

\appendixsection{Derivation of the ADE-corrected entropic stabilizer}
\label{app:ade-corrected-stabilizer}

This appendix derives the ghost amplitude used in the paper from the logarithmic expansion of the entropy-stationarity condition. The correction term $\mathcal{C}$ appears because the scalar ADE equilibrium contains velocity-dependent second-order kinetic content while the scalar transport model has no separate hydrodynamic stress sector.

\subsection{Entropy-stationarity condition}

After tensorial flux relaxation, but before the ghost relaxation, the intermediate state is
\begin{equation}
\widetilde{g}_i=g_i+\Omega_i^{(s)}
=g_i^{eq}+\Delta s_i^{\mathrm{post}}+\Delta h_i,
\end{equation}
where the superscript $\mathrm{post}$ denotes the flux-sector population after tensorial flux relaxation,
\begin{equation}
\Delta s_i^{\mathrm{post}}
=\mathcal{P}_i\!\left[(\mathbf{I}-\mathbf{S})\mathbf{j}\right].
\end{equation}
The ghost update gives
\begin{equation}
g_i^{\star}
=\widetilde{g}_i-\lambda\Delta h_i
=g_i^{eq}+\Delta s_i^{\mathrm{post}}+(1-\lambda)\Delta h_i.
\end{equation}
On the positive branch, the entropy is
\begin{equation}
H[g^{\star}]
=\sum_i g_i^{\star}\ln\left(\frac{g_i^{\star}}{W_i}\right).
\end{equation}
Differentiating in the ghost direction gives
\begin{equation}
\frac{dH[g^{\star}]}{d\lambda}
=-\sum_i\Delta h_i\left[
\ln\left(\frac{g_i^{\star}}{W_i}\right)+1
\right].
\end{equation}
Since $\sum_i\Delta h_i=0$, stationarity is equivalent to
\begin{equation}
\sum_i\Delta h_i
\ln\left(
\frac{g_i^{eq}+\Delta s_i^{\mathrm{post}}+(1-\lambda)\Delta h_i}{W_i}
\right)=0.
\label{eq:app-root}
\end{equation}

\subsection{Logarithmic expansion and closed-form amplitude}

Define
\begin{equation}
\delta_i(\lambda)
=\Delta s_i^{\mathrm{post}}+(1-\lambda)\Delta h_i.
\end{equation}
On the weakly non-equilibrium branch,
\begin{equation}
|\delta_i(\lambda)|\ll g_i^{eq},\qquad g_i^{eq}>0,
\end{equation}
so
\begin{equation}
\ln\left(
\frac{g_i^{eq}+\Delta s_i^{\mathrm{post}}+(1-\lambda)\Delta h_i}{W_i}
\right)
=\ln\left(\frac{g_i^{eq}}{W_i}\right)
+\frac{\delta_i(\lambda)}{g_i^{eq}}
+O\!\left(\frac{\delta_i(\lambda)^2}{(g_i^{eq})^2}\right).
\end{equation}
Substitution into \eqref{eq:app-root} gives, to first order,
\begin{equation}
\sum_i\Delta h_i\frac{\Delta s_i^{\mathrm{post}}}{g_i^{eq}}
+(1-\lambda)\sum_i\Delta h_i\frac{\Delta h_i}{g_i^{eq}}
+\sum_i\Delta h_i\ln\left(\frac{g_i^{eq}}{W_i}\right)=0.
\end{equation}
With
\begin{equation}
\langle x|y\rangle_H=\sum_i\frac{x_i y_i}{g_i^{eq}},
\qquad
\mathcal{C}=\sum_i\Delta h_i\ln\left(\frac{g_i^{eq}}{W_i}\right),
\end{equation}
this becomes
\begin{equation}
\langle\Delta h|\Delta s^{\mathrm{post}}\rangle_H
+(1-\lambda)\langle\Delta h|\Delta h\rangle_H
+\mathcal{C}=0.
\end{equation}
If $\Delta h_i\equiv0$, no ghost correction is needed and $\lambda=1$. Otherwise,
\begin{equation}
\lambda=1+
\frac{\langle\Delta h|\Delta s^{\mathrm{post}}\rangle_H+\mathcal{C}}
{\langle\Delta h|\Delta h\rangle_H}.
\label{eq:app-lambda}
\end{equation}
This is the ADE-corrected entropic amplitude used in the anisotropic formulation.

\subsection{Reduction to the isotropic scalar limit}

When the physical relaxation is isotropic, $\mathbf{S}=2\beta\mathbf{I}$ and
\begin{equation}
\Delta s_i^{\mathrm{post}}=(1-2\beta)\Delta s_i.
\end{equation}
With $\lambda=\beta\gamma$, \eqref{eq:app-lambda} gives
\begin{equation}
\beta\gamma
=1+\frac{(1-2\beta)\langle\Delta h|\Delta s\rangle_H+\mathcal{C}}
{\langle\Delta h|\Delta h\rangle_H},
\end{equation}
or
\begin{equation}
\gamma
=\frac{1}{\beta}
-\left(2-\frac{1}{\beta}\right)
\frac{\langle\Delta h|\Delta s\rangle_H}{\langle\Delta h|\Delta h\rangle_H}
+\frac{\mathcal{C}}{\beta\langle\Delta h|\Delta h\rangle_H}.
\end{equation}
This recovers the ADE-corrected scalar ELBM expression in the isotropic limit.

\subsection{Origin of the correction term}

The correction term is a consequence of the scalar ADE decomposition. In hydrodynamic KBC models for Navier--Stokes flow, the non-equilibrium stress sector is hydrodynamic, and the remaining ghost sector can be orthogonal to the equilibrium contribution at the relevant order. The passive-scalar ADE model has only one conserved scalar and a first-order transport flux. Its second-order kinetic content is therefore not a hydrodynamic stress mode, even though the equilibrium retains
\begin{equation}
\frac{(\mathbf{c}_i\cdot\mathbf{u})^2}{2c_s^4}
-\frac{\mathbf{u}\cdot\mathbf{u}}{2c_s^2}
\end{equation}
to control advective truncation errors. The overlap
\begin{equation}
\sum_i\Delta h_i\ln\left(\frac{g_i^{eq}}{W_i}\right)
\end{equation}
is not forced to vanish under this decomposition, so the explicit term $\mathcal{C}$ is kept in \eqref{eq:app-lambda}.

Anisotropy does not change the structure of this correction. The tensor $\mathbf{D}$ acts through the post-flux state
\begin{equation}
\Delta s_i^{\mathrm{post}}
=\mathcal{P}_i[(\mathbf{I}-\mathbf{S})\mathbf{j}],
\end{equation}
while $\mathcal{C}$ is tied to the scalar equilibrium and the ghost subspace. Thus the tensorial relaxation fixes the physical diffusive flux, and the ADE correction fixes the entropy balance of the higher-order kinetic sector used by the stabilizer.